
\font\twelverm=cmr10 scaled 1200    \font\twelvei=cmmi10 scaled 1200
\font\twelvesy=cmsy10 scaled 1200   \font\twelveex=cmex10 scaled 1200
\font\twelvebf=cmbx10 scaled 1200   \font\twelvesl=cmsl10 scaled 1200
\font\twelvett=cmtt10 scaled 1200   \font\twelveit=cmti10 scaled 1200

\skewchar\twelvei='177   \skewchar\twelvesy='60


\def\twelvepoint{\normalbaselineskip=12.4pt plus 0.1pt minus 0.1pt
  \abovedisplayskip 12.4pt plus 3pt minus 9pt
  \belowdisplayskip 12.4pt plus 3pt minus 9pt
  \abovedisplayshortskip 0pt plus 3pt
  \belowdisplayshortskip 7.2pt plus 3pt minus 4pt
  \smallskipamount=3.6pt plus1.2pt minus1.2pt
  \medskipamount=7.2pt plus2.4pt minus2.4pt
  \bigskipamount=14.4pt plus4.8pt minus4.8pt
  \def\rm{\fam0\twelverm}          \def\it{\fam\itfam\twelveit}%
  \def\sl{\fam\slfam\twelvesl}     \def\bf{\fam\bffam\twelvebf}%
  \def\mit{\fam 1}                 \def\cal{\fam 2}%
  \def\tt{\twelvett}
  \textfont0=\twelverm   \scriptfont0=\tenrm   \scriptscriptfont0=\sevenrm
  \textfont1=\twelvei    \scriptfont1=\teni    \scriptscriptfont1=\seveni
  \textfont2=\twelvesy   \scriptfont2=\tensy   \scriptscriptfont2=\sevensy
  \textfont3=\twelveex   \scriptfont3=\twelveex  \scriptscriptfont3=\twelveex
  \textfont\itfam=\twelveit
  \textfont\slfam=\twelvesl
  \textfont\bffam=\twelvebf \scriptfont\bffam=\tenbf
  \scriptscriptfont\bffam=\sevenbf
  \normalbaselines\rm}



\def\beginlinemode{\endmode
  \begingroup\parskip=0pt \obeylines\def\\{\par}\def\endmode{\par\endgroup}}
\def\beginparmode{\endmode
  \begingroup \def\endmode{\par\endgroup}}
\let\endmode=\par
{\obeylines\gdef\
{}}
\def\singlespace{\baselineskip=\normalbaselineskip}

\def\oneandahalfspace{\baselineskip=\normalbaselineskip
  \multiply\baselineskip by 3 \divide\baselineskip by 2}
\def\doublespace{\baselineskip=\normalbaselineskip \multiply\baselineskip by 2}

\newcount\firstpageno
\firstpageno=2
\footline=
{\ifnum\pageno<\firstpageno{\hfil}\else{\hfil\twelverm\folio\hfil}\fi}
\def\toppageno{\global\footline={\hfil}\global\headline
  ={\ifnum\pageno<\firstpageno{\hfil}\else{\hfil\twelverm\folio\hfil}\fi}}
\let\rawfootnote=\footnote		
\def\footnote#1#2{{\rm\singlespace\parindent=0pt\parskip=0pt
  \rawfootnote{#1}{#2\hfill\vrule height 0pt depth 6pt width 0pt}}}
\def\raggedcenter{\leftskip=4em plus 12em \rightskip=\leftskip
  \parindent=0pt \parfillskip=0pt \spaceskip=.3333em \xspaceskip=.5em
  \pretolerance=9999 \tolerance=9999
  \hyphenpenalty=9999 \exhyphenpenalty=9999 }
\def\dateline{\rightline{\ifcase\month\or
  January\or February\or March\or April\or May\or June\or
  July\or August\or September\or October\or November\or December\fi
  \space\number\year}}
\def\today{\ifcase\month\or
  January\or February\or March\or April\or May\or June\or
  July\or August\or September\or October\or November\or December\fi
  \space\number\day, \number\year}
\def\received{\vskip 3pt plus 0.2fill
 \centerline{\sl (Received\space\ifcase\month\or
  January\or February\or March\or April\or May\or June\or
  July\or August\or September\or October\or November\or December\fi
  \qquad, \number\year)}}


\hsize=6.5truein
\hoffset=0truein
\vsize=8.9truein
\voffset=0truein
\parskip=\medskipamount
\def\\{\cr}
\twelvepoint		
\doublespace		
\overfullrule=0pt	




\def\title			
  {\null\vskip 3pt plus 0.2fill
   \beginlinemode \doublespace \raggedcenter \bf}

\def\author			
  {\vskip 3pt plus 0.2fill \beginlinemode
   \singlespace \raggedcenter}

\def\affil			
  {\vskip 3pt plus 0.1fill \beginlinemode
   \oneandahalfspace \raggedcenter \sl}

\def\abstract			
  {\vskip 3pt plus 0.3fill \beginparmode
   \doublespace ABSTRACT: }

\def\submit  			
	{\vskip 24pt \beginlinemode
	\noindent \rm Submitted to: \sl}

\def\endtopmatter		
  {\endpage			
   \body}

\def\body			
  {\beginparmode}		

\def\head#1{			
  \goodbreak\vskip 0.5truein	
  {\immediate\write16{#1}
   \raggedcenter \uppercase{#1}\par}
   \nobreak\vskip 0.25truein\nobreak}

\def\beneathrel#1\under#2{\mathrel{\mathop{#2}\limits_{#1}}}

\def\refto#1{$^{#1}$}		

\def\references			
  {\head{References}		
   \beginparmode
   \frenchspacing \parindent=0pt \leftskip=1truecm
   \parskip=8pt plus 3pt \everypar{\hangindent=\parindent}}

\gdef\refis#1{\item{#1.\ }}			

\gdef\journal#1, #2, #3, 1#4#5#6{		
    {\sl #1~}{\bf #2}, #3 (1#4#5#6)}		

\def\endreferences{\body}

\def\figurecaptions		
  {\endpage
   \beginparmode
   \head{Figure Captions}
}

\def\endpage			
  {\vfill\eject}

\def\endpaper			
  {\endmode\vfill\supereject}

\def\endit
  {\endpaper\end}


\def\heading				
  {\vskip 0.5truein plus 0.1truein	
   \beginparmode \def\\{\par} \parskip=0pt \singlespace \raggedcenter}

\def\subheading				
  {\vskip 0.25truein plus 0.1truein	
   \beginlinemode \singlespace \parskip=0pt \def\\{\par}\raggedcenter}

\def\tag#1$${\eqno(#1)$$}

\def\align#1$${\eqalign{#1}$$}

\def\aligntag#1$${\gdef\tag##1\\{&(##1)\cr}\eqalignno{#1\\}$$
  \gdef\tag##1$${\eqno(##1)$$}}

\def\endaligntag{}

\def\overset#1\to#2{{\mathop{#2}^{#1}}}
\def\underset#1\to#2{{\mathop{#2}_{#1}}}

\def\enddocument{\endit}


\def\ref#1{Ref.~#1}			
\def\Ref#1{Ref.~#1}			
\def\[#1]{[\cite{#1}]}
\def\cite#1{{#1}}
\def\(#1){(\call{#1})}
\def\call#1{{#1}}
\def\taghead#1{}
\def\frac#1#2{{#1 \over #2}}

\def\12{{1\over2}}

\def\sla{\raise.15ex\hbox{$/$}\kern-.57em}
\def\leaderfill{\leaders\hbox to 1em{\hss.\hss}\hfill}
\def\twiddle{\lower.9ex\rlap{$\kern-.1em\scriptstyle\sim$}}
\def\bigtwiddle{\lower1.ex\rlap{$\sim$}}
\def\gtwid{\mathrel{\raise.3ex\hbox{$>$\kern-.75em\lower1ex\hbox{$\sim$}}}}
\def\ltwid{\mathrel{\raise.3ex\hbox{$<$\kern-.75em\lower1ex\hbox{$\sim$}}}}
\def\square{\kern1pt\vbox{\hrule height 1.2pt\hbox{\vrule width 1.2pt\hskip 3pt
   \vbox{\vskip 6pt}\hskip 3pt\vrule width 0.6pt}\hrule height 0.6pt}\kern1pt}
\def\tdot#1{\mathord{\mathop{#1}\limits^{\kern2pt\ldots}}}

\def\pmb#1{\setbox0=\hbox{#1}%
  \kern-.025em\copy0\kern-\wd0
  \kern  .05em\copy0\kern-\wd0
  \kern-.025em\raise.0433em\box0 }

\catcode`@=11
\newcount\r@fcount \r@fcount=0
\newcount\r@fcurr
\immediate\newwrite\reffile
\newif\ifr@ffile\r@ffilefalse
\def\w@rnwrite#1{\ifr@ffile\immediate\write\reffile{#1}\fi\message{#1}}

\def\writer@f#1>>{}
\def\referencefile{
  \r@ffiletrue\immediate\openout\reffile=\jobname.ref%
  \def\writer@f##1>>{\ifr@ffile\immediate\write\reffile%
    {\noexpand\refis{##1} = \csname r@fnum##1\endcsname = %
     \expandafter\expandafter\expandafter\strip@t\expandafter%
     \meaning\csname r@ftext\csname r@fnum##1\endcsname\endcsname}\fi}%
  \def\strip@t##1>>{}}

\def\citeall#1{\xdef#1##1{#1{\noexpand\cite{##1}}}}
\def\cite#1{\each@rg\citer@nge{#1}}	

\def\each@rg#1#2{{\let\thecsname=#1\expandafter\first@rg#2,\end,}}
\def\first@rg#1,{\thecsname{#1}\apply@rg}	
\def\apply@rg#1,{\ifx\end#1\let\next=\relax
\else,\thecsname{#1}\let\next=\apply@rg\fi\next}

\def\citer@nge#1{\citedor@nge#1-\end-}	
\def\citer@ngeat#1\end-{#1}
\def\citedor@nge#1-#2-{\ifx\end#2\r@featspace#1 
  \else\citel@@p{#1}{#2}\citer@ngeat\fi}	
\def\citel@@p#1#2{\ifnum#1>#2{\errmessage{Reference range #1-#2\space is bad.}
    \errhelp{If you cite a series of references by the notation M-N, then M and
    N must be integers, and N must be greater than or equal to M.}}\else%
 {\count0=#1\count1=#2\advance\count1
by1\relax\expandafter\r@fcite\the\count0,%
  \loop\advance\count0 by1\relax
    \ifnum\count0<\count1,\expandafter\r@fcite\the\count0,%
  \repeat}\fi}

\def\r@featspace#1#2 {\r@fcite#1#2,}	
\def\r@fcite#1,{\ifuncit@d{#1}		
    \expandafter\gdef\csname r@ftext\number\r@fcount\endcsname%
    {\message{Reference #1 to be supplied.}\writer@f#1>>#1 to be supplied.\par
     }\fi%
  \csname r@fnum#1\endcsname}

\def\ifuncit@d#1{\expandafter\ifx\csname r@fnum#1\endcsname\relax%
\global\advance\r@fcount by1%
\expandafter\xdef\csname r@fnum#1\endcsname{\number\r@fcount}}

\let\r@fis=\refis			
\def\refis#1#2#3\par{\ifuncit@d{#1}
    \w@rnwrite{Reference #1=\number\r@fcount\space is not cited up to now.}\fi%
  \expandafter\gdef\csname r@ftext\csname r@fnum#1\endcsname\endcsname%
  {\writer@f#1>>#2#3\par}}

\def\r@ferr{\endreferences\errmessage{I was expecting to see
\noexpand\endreferences before now;  I have inserted it here.}}
\let\r@ferences=\references
\def\references{\r@ferences\def\endmode{\r@ferr\par\endgroup}}

\let\endr@ferences=\endreferences
\def\endreferences{\r@fcurr=0
  {\loop\ifnum\r@fcurr<\r@fcount
    \advance\r@fcurr by 1\relax\expandafter\r@fis\expandafter{\number\r@fcurr}%
    \csname r@ftext\number\r@fcurr\endcsname%
  \repeat}\gdef\r@ferr{}\endr@ferences}


\let\r@fend=\endpaper\gdef\endpaper{\ifr@ffile
\immediate\write16{Cross References written on []\jobname.REF.}\fi\r@fend}

\catcode`@=12

\citeall\refto		
\citeall\ref		%
\citeall\Ref		%

\catcode`@=11
\newcount\tagnumber\tagnumber=0

\immediate\newwrite\eqnfile
\newif\if@qnfile\@qnfilefalse
\def\write@qn#1{}
\def\writenew@qn#1{}
\def\w@rnwrite#1{\write@qn{#1}\message{#1}}
\def\@rrwrite#1{\write@qn{#1}\errmessage{#1}}

\def\taghead#1{\gdef\t@ghead{#1}\global\tagnumber=0}
\def\t@ghead{}

\expandafter\def\csname @qnnum-3\endcsname
  {{\t@ghead\advance\tagnumber by -3\relax\number\tagnumber}}
\expandafter\def\csname @qnnum-2\endcsname
  {{\t@ghead\advance\tagnumber by -2\relax\number\tagnumber}}
\expandafter\def\csname @qnnum-1\endcsname
  {{\t@ghead\advance\tagnumber by -1\relax\number\tagnumber}}
\expandafter\def\csname @qnnum0\endcsname
  {\t@ghead\number\tagnumber}
\expandafter\def\csname @qnnum+1\endcsname
  {{\t@ghead\advance\tagnumber by 1\relax\number\tagnumber}}
\expandafter\def\csname @qnnum+2\endcsname
  {{\t@ghead\advance\tagnumber by 2\relax\number\tagnumber}}
\expandafter\def\csname @qnnum+3\endcsname
  {{\t@ghead\advance\tagnumber by 3\relax\number\tagnumber}}

\def\equationfile{%
  \@qnfiletrue\immediate\openout\eqnfile=\jobname.eqn%
  \def\write@qn##1{\if@qnfile\immediate\write\eqnfile{##1}\fi}
  \def\writenew@qn##1{\if@qnfile\immediate\write\eqnfile
    {\noexpand\tag{##1} = (\t@ghead\number\tagnumber)}\fi}
}

\def\callall#1{\xdef#1##1{#1{\noexpand\call{##1}}}}
\def\call#1{\each@rg\callr@nge{#1}}

\def\each@rg#1#2{{\let\thecsname=#1\expandafter\first@rg#2,\end,}}
\def\first@rg#1,{\thecsname{#1}\apply@rg}
\def\apply@rg#1,{\ifx\end#1\let\next=\relax%
\else,\thecsname{#1}\let\next=\apply@rg\fi\next}

\def\callr@nge#1{\calldor@nge#1-\end-}
\def\callr@ngeat#1\end-{#1}
\def\calldor@nge#1-#2-{\ifx\end#2\@qneatspace#1 %
  \else\calll@@p{#1}{#2}\callr@ngeat\fi}
\def\calll@@p#1#2{\ifnum#1>#2{\@rrwrite{Equation range #1-#2\space is bad.}
\errhelp{If you call a series of equations by the notation M-N, then M and
N must be integers, and N must be greater than or equal to M.}}\else%
 {\count0=#1\count1=#2\advance\count1
by1\relax\expandafter\@qncall\the\count0,%
  \loop\advance\count0 by1\relax%
    \ifnum\count0<\count1,\expandafter\@qncall\the\count0,%
  \repeat}\fi}

\def\@qneatspace#1#2 {\@qncall#1#2,}
\def\@qncall#1,{\ifunc@lled{#1}{\def\next{#1}\ifx\next\empty\else
  \w@rnwrite{Equation number \noexpand\(>>#1<<) has not been defined yet.}
  >>#1<<\fi}\else\csname @qnnum#1\endcsname\fi}

\let\eqnono=\eqno
\def\eqno(#1){\tag#1}
\def\tag#1$${\eqnono(\displayt@g#1 )$$}

\def\aligntag#1\endaligntag
  $${\gdef\tag##1\\{&(##1 )\cr}\eqalignno{#1\\}$$
  \gdef\tag##1$${\eqnono(\displayt@g##1 )$$}}

\def\eqalignno#1{\displ@y \tabskip\centering
  \halign to\displaywidth{\hfil$\displaystyle{##}$\tabskip\z@skip
    &$\displaystyle{{}##}$\hfil\tabskip\centering
    &\llap{$\displayt@gpar##$}\tabskip\z@skip\crcr
    #1\crcr}}

\def\displayt@gpar(#1){(\displayt@g#1 )}

\def\displayt@g#1 {\rm\ifunc@lled{#1}\global\advance\tagnumber by1
        {\def\next{#1}\ifx\next\empty\else\expandafter
        \xdef\csname @qnnum#1\endcsname{\t@ghead\number\tagnumber}\fi}%
  \writenew@qn{#1}\t@ghead\number\tagnumber\else
        {\edef\next{\t@ghead\number\tagnumber}%
        \expandafter\ifx\csname @qnnum#1\endcsname\next\else
        \w@rnwrite{Equation \noexpand\tag{#1} is a duplicate number.}\fi}%
  \csname @qnnum#1\endcsname\fi}

\def\ifunc@lled#1{\expandafter\ifx\csname @qnnum#1\endcsname\relax}

\let\@qnend=\end\gdef\end{\if@qnfile
\immediate\write16{Equation numbers written on []\jobname.EQN.}\fi\@qnend}

\catcode`@=12


%
%
%
%
%
\catcode`\@=11\relax
\newwrite\@unused
\def\typeout#1{{\let\protect\string\immediate\write\@unused{#1}}}
\typeout{psfig: version 1.1}

%
%
\def\@nnil{\@nil}
\def\@empty{}
\def\@psdonoop#1\@@#2#3{}
\def\@psdo#1:=#2\do#3{\edef\@psdotmp{#2}\ifx\@psdotmp\@empty \else
    \expandafter\@psdoloop#2,\@nil,\@nil\@@#1{#3}\fi}
\def\@psdoloop#1,#2,#3\@@#4#5{\def#4{#1}\ifx #4\@nnil \else
       #5\def#4{#2}\ifx #4\@nnil \else#5\@ipsdoloop #3\@@#4{#5}\fi\fi}
\def\@ipsdoloop#1,#2\@@#3#4{\def#3{#1}\ifx #3\@nnil
       \let\@nextwhile=\@psdonoop \else
      #4\relax\let\@nextwhile=\@ipsdoloop\fi\@nextwhile#2\@@#3{#4}}
\def\@tpsdo#1:=#2\do#3{\xdef\@psdotmp{#2}\ifx\@psdotmp\@empty \else
    \@tpsdoloop#2\@nil\@nil\@@#1{#3}\fi}
\def\@tpsdoloop#1#2\@@#3#4{\def#3{#1}\ifx #3\@nnil
       \let\@nextwhile=\@psdonoop \else
      #4\relax\let\@nextwhile=\@tpsdoloop\fi\@nextwhile#2\@@#3{#4}}
\def\psdraft{
	\def\@psdraft{0}
}
\def\psfull{
	\def\@psdraft{100}
}
\psfull
\newif\if@prologfile
\newif\if@postlogfile
\newif\if@bbllx
\newif\if@bblly
\newif\if@bburx
\newif\if@bbury
\newif\if@height
\newif\if@width
\newif\if@rheight
\newif\if@rwidth
\newif\if@clip
\def\@p@@sclip#1{\@cliptrue}
\def\@p@@sfile#1{
		   \def\@p@sfile{#1}
}
\def\@p@@sfigure#1{\def\@p@sfile{#1}}
\def\@p@@sbbllx#1{
		\@bbllxtrue
		\dimen100=#1
		\edef\@p@sbbllx{\number\dimen100}
}
\def\@p@@sbblly#1{
		\@bbllytrue
		\dimen100=#1
		\edef\@p@sbblly{\number\dimen100}
}
\def\@p@@sbburx#1{
		\@bburxtrue
		\dimen100=#1
		\edef\@p@sbburx{\number\dimen100}
}
\def\@p@@sbbury#1{
		\@bburytrue
		\dimen100=#1
		\edef\@p@sbbury{\number\dimen100}
}
\def\@p@@sheight#1{
		\@heighttrue
		\dimen100=#1
   		\edef\@p@sheight{\number\dimen100}
}
\def\@p@@swidth#1{
		\@widthtrue
		\dimen100=#1
		\edef\@p@swidth{\number\dimen100}
}
\def\@p@@srheight#1{
		\@rheighttrue
		\dimen100=#1
		\edef\@p@srheight{\number\dimen100}
}
\def\@p@@srwidth#1{
		\@rwidthtrue
		\dimen100=#1
		\edef\@p@srwidth{\number\dimen100}
}
\def\@p@@sprolog#1{\@prologfiletrue\def\@prologfileval{#1}}
\def\@p@@spostlog#1{\@postlogfiletrue\def\@postlogfileval{#1}}
\def\@cs@name#1{\csname #1\endcsname}
\def\@setparms#1=#2,{\@cs@name{@p@@s#1}{#2}}
%
%
\def\ps@init@parms{
		\@bbllxfalse \@bbllyfalse
		\@bburxfalse \@bburyfalse
		\@heightfalse \@widthfalse
		\@rheightfalse \@rwidthfalse
		\def\@p@sbbllx{}\def\@p@sbblly{}
		\def\@p@sbburx{}\def\@p@sbbury{}
		\def\@p@sheight{}\def\@p@swidth{}
		\def\@p@srheight{}\def\@p@srwidth{}
		\def\@p@sfile{}
		\def\@p@scost{10}
		\def\@sc{}
		\@prologfilefalse
		\@postlogfilefalse
		\@clipfalse
}
%
%
\def\parse@ps@parms#1{
	 	\@psdo\@psfiga:=#1\do
		   {\expandafter\@setparms\@psfiga,}}
%
%
\newif\ifno@bb
\newif\ifnot@eof
\newread\ps@stream
\def\bb@missing{
	\typeout{psfig: searching \@p@sfile \space  for bounding box}
	\openin\ps@stream=\@p@sfile
	\no@bbtrue
	\not@eoftrue
	\catcode`\%=12
	\loop
		\read\ps@stream to \line@in
		\global\toks200=\expandafter{\line@in}
		\ifeof\ps@stream \not@eoffalse \fi
		\@bbtest{\toks200}
		\if@bbmatch\not@eoffalse\expandafter\bb@cull\the\toks200\fi
	\ifnot@eof \repeat
	\catcode`\%=14
}
\catcode`\%=12
\newif\if@bbmatch
\def\@bbtest#1{\expandafter\@a@\the#1
\long\def\@a@#1
{\ifx\@bbtest#2\@bbmatchfalse\else\@bbmatchtrue\fi}
\long\def\bb@cull#1 #2 #3 #4 #5 {
	\dimen100=#2 bp\edef\@p@sbbllx{\number\dimen100}
	\dimen100=#3 bp\edef\@p@sbblly{\number\dimen100}
	\dimen100=#4 bp\edef\@p@sbburx{\number\dimen100}
	\dimen100=#5 bp\edef\@p@sbbury{\number\dimen100}
	\no@bbfalse
}
\catcode`\%=14
\def\compute@bb{
		\no@bbfalse
		\if@bbllx \else \no@bbtrue \fi
		\if@bblly \else \no@bbtrue \fi
		\if@bburx \else \no@bbtrue \fi
		\if@bbury \else \no@bbtrue \fi
		\ifno@bb \bb@missing \fi
		\ifno@bb \typeout{FATAL ERROR: no bb supplied or found}
			\no-bb-error
		\fi
		\count203=\@p@sbburx
		\count204=\@p@sbbury
		\advance\count203 by -\@p@sbbllx
		\advance\count204 by -\@p@sbblly
		\edef\@bbw{\number\count203}
		\edef\@bbh{\number\count204}
}
%
%
\def\in@hundreds#1#2#3{\count240=#2 \count241=#3
		     \count100=\count240	
		     \divide\count100 by \count241
		     \count101=\count100
		     \multiply\count101 by \count241
		     \advance\count240 by -\count101
		     \multiply\count240 by 10
		     \count101=\count240	
		     \divide\count101 by \count241
		     \count102=\count101
		     \multiply\count102 by \count241
		     \advance\count240 by -\count102
		     \multiply\count240 by 10
		     \count102=\count240	
		     \divide\count102 by \count241
		     \count200=#1\count205=0
		     \count201=\count200
			\multiply\count201 by \count100
		 	\advance\count205 by \count201
		     \count201=\count200
			\divide\count201 by 10
			\multiply\count201 by \count101
			\advance\count205 by \count201
		     \count201=\count200
			\divide\count201 by 100
			\multiply\count201 by \count102
			\advance\count205 by \count201
		     \edef\@result{\number\count205}
}
\def\compute@wfromh{
		\in@hundreds{\@p@sheight}{\@bbw}{\@bbh}
		\edef\@p@swidth{\@result}
}
\def\compute@hfromw{
		\in@hundreds{\@p@swidth}{\@bbh}{\@bbw}
		\edef\@p@sheight{\@result}
}
\def\compute@handw{
		\if@height
			\if@width
			\else
				\compute@wfromh
			\fi
		\else
			\if@width
				\compute@hfromw
			\else
				\edef\@p@sheight{\@bbh}
				\edef\@p@swidth{\@bbw}
			\fi
		\fi
}
\def\compute@resv{
		\if@rheight \else \edef\@p@srheight{\@p@sheight} \fi
		\if@rwidth \else \edef\@p@srwidth{\@p@swidth} \fi
}
%
\def\compute@sizes{
	\compute@bb
	\compute@handw
	\compute@resv
}
%
%
\def\psfig#1{\vbox {
	%
	\ps@init@parms
	\parse@ps@parms{#1}
	\compute@sizes
	\ifnum\@p@scost<\@psdraft{
		\typeout{psfig: including \@p@sfile \space }
		\special{ps::[begin] 	\@p@swidth \space \@p@sheight \space
				\@p@sbbllx \space \@p@sbblly \space
				\@p@sbburx \space \@p@sbbury \space
				startTexFig \space }
		\if@clip{
			\typeout{(clip)}
			\special{ps:: \@p@sbbllx \space \@p@sbblly \space
				\@p@sbburx \space \@p@sbbury \space
				doclip \space }
		}\fi
		\if@prologfile
		    \special{ps: plotfile \@prologfileval \space } \fi
		\special{ps: plotfile \@p@sfile \space }
		\if@postlogfile
		    \special{ps: plotfile \@postlogfileval \space } \fi
		\special{ps::[end] endTexFig \space }
		\vbox to \@p@srheight true sp{
			\hbox to \@p@srwidth true sp{
				\hfil
			}
		\vfil
		}
	}\else{
		\vbox to \@p@srheight true sp{
		\vss
			\hbox to \@p@srwidth true sp{
				\hss
				\@p@sfile
				\hss
			}
		\vss
		}
	}\fi
}}
\catcode`\@=12\relax


\xdef\picsize{4.0truein}
\def \com#1{}
\def\figdira{figuredir/figdirafakes/}
\def\figdira{figuredir/figdira/}

\def\figdirc{figuredir/figdircfakes/}
\def\figdirc{figuredir/figdirc/}

\def\figdird{figuredir/figdirdfakes/}
\def\figdird{figuredir/figdird/}

\title  Disorder-Driven Pretransitional Tweed in Martensitic Transformations
\author Sivan Kartha
\affil Institute for Advanced Study, Princeton, New Jersey 08540
\author James~A. Krumhansl\footnote{$^\dagger$}{Present Address: 515 Station
Rd., Amherst, Massachusetts 01002}, James~P. Sethna, and L. K. Wickham
\affil Laboratory of Atomic and Solid State Physics, Cornell University, %
Ithaca, New York 14853-2501

\abstract
Defying the conventional wisdom regarding first--order transitions,
{\it solid--solid displacive transformations} are often accompanied by
pronounced pretransitional phenomena.  Generally, these phenomena are
indicative of some mesoscopic lattice deformation that ``anticipates''
the upcoming phase transition.  Among these precursive effects is the
observation of the so-called ``tweed'' pattern in transmission
electron microscopy in a wide variety of materials. We have
investigated the tweed deformation in a two dimensional model system,
and found that it arises because the compositional disorder intrinsic
to any alloy conspires with the natural geometric constraints of the
lattice to produce a frustrated, glassy phase.  The predicted phase
diagram and glassy behavior have been verified by numerical
simulations, and diffraction patterns of simulated systems are found
to compare well with experimental data.  Analytically comparing to
alternative models of strain-disorder coupling, we show that the
present model best accounts for experimental observations.

\noindent PACS numbers: 81.30.Kf, 75.10.Nr, 61.70.Wp

\endtopmatter

\noindent{\bf Introduction}\hfil\break
Typically, first--order transformations occur abruptly.  The
liquid--vapor phase change, for example, is not heralded by critical
fluctuations, length scales do not start diverging, and the system
does not demonstrate large anticipatory excursions into the
approaching phase.  The textbook first--order behavior is fairly
uneventful compared to universal critical phenomena associated with
second--order transitions.  In marked contrast to this
well--established pattern, first-order {\it solid-solid} structural
transformations (e.g. {\it martensitic} transformations) demonstrate
pretransitional effects for as much as {\it hundreds} of degrees above
the nominal transition temperature, despite their distinctly
first--order nature.  As witnessed in a wide--ranging variety of
martensitic materials, this striking pretransitional behavior takes
several different forms: anomalous phenomena in x-ray, electron, and
neutron scattering including the quasi-elastic ``central peak''
observation in neutron scattering; partial elastic softening of
various lattice distortive modes, including $q = 0$ homogeneous
deformations as well as $q \ne 0$ phonon modes; anomalous behavior in
transport coefficients and thermal expansion coefficients.  One
particularly distinctive example of such precursor phenomena is the
observation of the ``tweed'' pattern (Figure~1) in transmission
electron microscope images of materials approaching their martensitic
transformation\refto{TannersPicture}.  In this paper we study tweed in
materials undergoing martensitic transformations, with the aim of
better understanding the origin and nature of pretransitional
phenomena.

The main finding is that {\it disorder}, which is known to generally
be important in these materials, may in fact play a fundamental role
in bringing about pretransitional behavior, and that tweed can be
generated as a direct response even to the simple statistical
compositional disorder which is unavoidable in alloys.  (Special
defects are {\it not} required.)  Further, we provide numerical
evidence for the glassy behavior of tweed which earlier analysis had
predicted\refto{Kartha}.  Our approach is 1) to introduce a model
which exhibits a disorder-driven precursive tweed structure and to
detail its phase diagram, 2) to analyze the simulated x-ray
diffraction data and real space images, and 3) to analytically compare
this and other strain-disorder couplings by equating the
disorder-induced long--range elastic forces with a non-local
interaction in the order parameter, showing that the present model
best accounts for experimental observations.  The central importance
of a disorder in any model which hopes to shed light on
pretransitional phenomena is suggested by other experimental and
computational studies as well, including the findings of Petry {\it et
al}.\refto{Petry}, who observe precursor effects in zirconium doped
with small amounts of oxygen but not in pure zirconium, and Becquart
{\it et al.}\refto{Becquart}, who observe tweed structures in
molecular dynamics simulations of disordered materials but not ordered
materials.

\noindent{\bf Background}\hfil\break
Many materials of technological importance undergo martensitic
transformations.  In these solid-solid first-order structural
transformations, the lattice deforms from one crystalline structure to
another through some large-scale motion that preserves the topological
integrity of the lattice, (i.e. there exists a ``lattice
correspondence.'')  Unlike diffusive or order--disorder
transformations requiring the interchange of atoms, these
transformations are not {\it reconstructive}\refto{TannerWuttig}, that
is, bonds between neighbors are not broken and re-formed; there is no
diffusion and atoms maintain their relationship with their neighbors.
Rather, these transformations are {\it displacive}, meaning there is
some homogeneous strain that transforms one lattice into the other,
with atoms moving in a cooperative fashion, sometimes at sonic speeds.
A simple example is the transformation of a square lattice into a
rectangular lattice, brought about by stretching along one axis and
shrinking along the other.  Such transformations have sometimes been
termed ``military '' transformations, in order to convey the
impression of a large-scale coordinated motion of an entire lattice,
proceeding in lock--step from one configuration to another.  This is
in contrast to the relative anarchy of, say, a diffusive
transformation, in which the atoms wander in search of a locally
favorable environment, e.g.  molecules in a vapor diffusing to find a
home on a droplet.

The tweed pattern is seen as a pretransitional effect in transmission
electron microscopy of many different materials, including shape
memory alloys (NiAl\refto{Tweed_NiAl}, FePd\refto{Tweed_FePd,
Oshima_Hysteresis, Muto_Softening}, CuAu\refto{Tweed_CuAu} etc.)  and
high temperature superconductors (A-15 compounds\refto{Tweed_A15s}
V$_3$Si, Nb$_3$Sn, and the very high T$_c$
YBaCuO--type\refto{Tweed_YBCO, Zhu, Moss} and LaCuO--type cuprates
etc), various other ceramics\refto{Heuer} (e.g.  Y$_2$O$_3$-ZrO$_2$),
and alloys undergoing phase separation\refto{Tweed_CuBe} such as steel
during tempering treatment\refto{Tweed_steel}.  As suggested by its
name, the tweed pattern consists of diagonal striations bearing a
striking resemblance to the tweed textile.  The image has no strict
periodicity, but there are two apparent length scales: one (call it
$L$) corresponding to the longitudinal extent of the long diagonal
striations and the other to their relatively short transverse width
($\xi$).  These distances {\it appear} to be on the scale of tens or
hundreds of lattice constants.  However, it can be difficult to
determine the lengthscales unambiguously, as artifacts of the imaging
process can sometimes be confounded with genuine effects of the atomic
configuration.


In order both to develop a formal theory and to test it by simulation,
we consider as a model system a two dimensional solid which undergoes
a structural phase transformation from a square lattice to a
rectangular lattice as temperature is lowered\refto{Khachaturyan,
Parlinski, Falk, Jacobs}.  The two dimensional square $\rightarrow$
rectangular transition corresponds to the tetragonal $\rightarrow$
orthorhombic transition seen in planar compounds such as the
YBaCuO-type and LaCuO-type high--T$_c$ superconducting oxides.
Conceptually, this is also the two dimensional analog of the cubic
$\rightarrow$ tetragonal transition seen in many materials, such as
certain ferrous steels, shape memory alloys such as FePd and certain
Indium alloys, and the superconducting A-15 compounds Nb$_3$Sn and
V$_3$Si.

One very general and important experimental observation has attracted
our attention to the role of disorder in these systems.  Typically,
alloys undergoing martensitic transformations are extremely sensitive
to the relative alloying percentages of the elements which make them
up.  For example, Fe$_{1-\eta}$Pd$_\eta$ undergoes its martensitic
transformation at room temperature when $\eta = 29\%$, but as $\eta$
is increased to $32\%$, the transformation temperature, $T_M$,
plummets to absolute zero: a one percent shift in the concentration of
palladium causes a drastic $100^{\circ}K$ drop in $T_M$\ !  A
convenient way to think of this drastic compositional sensitivity is
to consider the martensitic transformation as a fixed temperature
phase transition which occurs as $\eta$ is varied and passes through
some critical composition.  The drastic dependence of transformation
temperature on composition can then be viewed as simply a weak
temperature dependence of the critical composition.

This drastic composition dependence of the transformation temperature
is a commonly observed property of many of the martensitic materials
that show pretransitional behavior.  In our opinion, this striking
property bears directly on the question of precursors and it offers
some insight into their ubiquity.  Since composition in any alloy or
doped compound is a spatially inhomogeneous quantity, the {\it actual}
composition will vary around some {\it average} composition simply due
to the disorder that is frozen in as the solid crystallizes from the melt.
Since the transformation temperature is so sensitive to composition,
there must exist a locally defined hypothetical transformation
temperature which depends on the local composition\refto{Imry}.  This
local transformation temperature may be higher or lower than the
observed transformation temperature, at which the first sign of bulk
transformation is observed in a given sample, and long range
martensitic order is actually established.  For example, a small
region in a sample of FdPd which has a lower than average
concentration of palladium will seek to transform into the martensitic
phase well before the transformation temperature at which the bulk
martensitic order actually develops.  The static, quenched--in, purely
statistical compositional disorder will determine the spatial
variation of local transformation temperature, and will thereby lead
to pretransitional deformations occurring on a mesoscopic scale in a
otherwise untransformed lattice.

In actuality any local tendency to transform may be suppressed by the
surroundings which may not be ready to transform.  Therefore, the
essence of this problem is to treat the overall system as a collection
of local regions which {\it interact} via extended strain fields.  It
cannot be a simple superpostion of different transformable units, nor
can models which simply address isolated defects produce the
pretransitional effects we propose.  Cooperative behavior, we believe,
is of the essence.

\noindent{\bf Theory and Model}\hfil\break
We seek here to model this coupling between compositional disorder and
the martensitic transformation, and to understand the nature of the
lattice deformation which arises in response.  Since the tweed
structure we are investigating is a lattice deformation with a length
scale of many lattice constants, we will adopt a perspective which
focuses on this mesoscale structure, and leaves the atomistic behavior
of the material largely unspecified.  To this end, we will view the
material as an elastic continuum, and analyze it within a
Landau-Ginzburg framework governing the lattice distortive free
energy.  Pursuing this approach, we construct a general free energy
which is consistent with the symmetries of the system and which is
taken to sufficiently high order in the relevant strain order
parameters to produce the important physical behavior.  The
parameters in the resulting free energy are related to empirically
measurable materials constants, such as elastic constants, phonon
dispersion curves, couplings to impurities, lattice constants etc.

At the outset, we emphasize that the fundamental cause of tweed in our
theory is simply local (static) variations in the effective coarse
grained free energy arising from compositional variation.  While
models which also couple to reconstructive ordering, or other chemical
reactions, have yielded tweed in simulations\refto{Khachaturyan,
Parlinski}, we suggest that ordering, {\it per se}, is not a
fundamental cause of tweed.  No ordering or reconstruction of any kind
takes place in the tweed and martensite regimes of almost any of the
well known alloys or ceramics that show this precursor behavior.
Perhaps, in fact, it is best to think of ordering or other replacive
effects simply as additional ways, beyond composition, to produce
spatial variation in the coarse grained free energy of our model.
Again, then, it is the cooperative elastic behavior of regions of
locally different coarse grained free energy which we propose to be
the generic origin of this kind of precursor.

The two dimensional system is modeled by the following free energy
relative to a perfect square reference lattice:
$$f = {{A_1} \over 2} e_1^2 + {A_2\over 2} e_2^2 + {A_\phi\over 2} \phi^2 -
{\beta\over 4} \phi^4 + {\gamma\over 6} \phi^6 + {\kappa\over 2} (\nabla
\phi)^2\eqno(free)$$ which is a functional of the strain fields
$e_1(x,y)$, $e_2(x,y)$, and $\phi(x,y)$.  Here, $e_1
\equiv (e_{xx} + e_{yy})/\sqrt{2}$ is the bulk dilational strain;
$e_2 \equiv e_{xy}$ is the shear strain; and $\phi \equiv (e_{xx} -
e_{yy})/\sqrt{2}$ is the deviatoric (or rectangular)
strain\refto{properferro}.  The symmetric strain tensor
${\bf e}$ is defined in the standard way\refto{Landau} as the
non-rotational part of the displacement gradients, $$e_{ij} \equiv {1
\over 2} ({\partial U_i\over\partial x_j} + {\partial U_j \over
\partial x_i} + {\partial U_l \over
\partial x_i} {\partial U_l \over \partial x_j}).
\eqno(straintensor)$$ The second order term guarantees that
finite rotations are not included in the strain tensor, but in general
this term is very small for these applications, and we have safely
neglected it in our analytical work, although all numerical
simulations include it.

The first three terms in the free energy \(free) simply mean that the
material in question has a Hooke's law restoring force to deformations
into the dilational, shear, and rectangular strain modes.  (These,
of course, are the only three homogeneous elastic modes available to a
two dimensional solid with square symmetry.)  The coefficients in
front of those three terms are simply the harmonic elastic constants
of the material, where $A_1 = C_{11} + C_{12}$, $A_2 = 4C_{44}$, and
$A_\phi = C_{11} - C_{12} = 2 C'$.  The free energy also includes
higher order (i.e.  anharmonic) terms in the rectangular strain,
$\phi$, in order to produce a first--order phase transition from a
square ``austenite'' phase with $\phi = 0$ to a rectangular
``martensite'' phase with $\phi = \pm \phi_M$.  This phase transition
occurs as the elastic constant $A_\phi$ ``softens'', (i.e.  decreases
with temperature), and below a (nonzero!)  critical value $A_\phi^{crit}
= {3\beta^2 \over 16 \gamma}$ the rectangular phase becomes the stable
phase, with the transformation strain $\phi_M = ((\beta +
\sqrt{\beta^2 - 4 A_\phi \gamma})/2\gamma)^{1\over2}$.  The parameters
$\beta$ and $\gamma$ are determined by the magnitude and the energy of
the martensitic strain at the critical
temperature\refto{partialsoftening}

We introduce compositional disorder in the simplest way which is
consistent with the symmetries of the problem.  The elastic constant
$A_\phi$ is allowed to be not only temperature dependent, but also
composition dependent.  The dependence of $A_\phi$ on both temperature and
composition is taken to be a simple linear relationship, thereby
quadratically coupling the strain order parameter $\phi$ to the random
composition field:
$$A_\phi({\bf x}) = A_T \cdot (T-T_0(\bar\eta)) +
A_\eta \cdot \delta\eta ({\bf x})
\equiv \bar{A_\phi} + \delta A_\phi.\eqno(couplingterm)$$
Here, $T_0(\bar\eta)$ is the temperature marking the mechanical
instability of the austenite phase at the nominal composition
$\bar\eta$, and $A_T$ and $A_\eta$ describe the linear dependence of
the elastic constant $A_\phi$ on temperature and composition,
respectively.  The spatially varying (but temporally constant) field
$\delta\eta ({\bf x}) \equiv \eta({\bf x}) - \bar\eta$ is the local
deviation from the average composition.  Its value on each simulation
cell is determined by selecting a random value from a gaussian
distribution of unit width.  (By normalizing to unit width, the {\it
magnitude} of compositional inhomogeneity and the {\it strength} of
its coupling to the elastic constant $A_\phi$ are both included in the
coupling parameter $A_\eta$.)  Then, the local ``transformation
temperature'' is given by $T_M({\bf x}) = T_M(\overline{\eta}) -
{A_\eta \over {A_T}} \cdot \delta\eta ({\bf x})$.

At high temperatures, all the regions of the system will prefer to be
in the undeformed phase, and at low temperatures all will prefer the
martensitic phase.  However, near the bulk transformation temperature,
there will be a temperature range given by the typical magnitude of
${A_\eta \over {A_T}} \cdot \delta\eta ({\bf x}) $, where the coupling
between the strain $\phi$ and the random compositional disorder can
provide a non-negligible driving force toward a pretransitional
deformation.  Using numerical simulations and analysis, we shall show
that a pretransitional deformation does occur for the present model,
and that it is tweed.

\noindent{\bf Simulation}\hfil\break
Figure 2 displays the results of a computer simulation of the model
described above.  Configurations 2(a-e) show the development from the
undeformed austenite phase, through a pretransitional regime, into the
fully developed martensite phase, as the elastic constant $A_\phi$
softens.  One immediately recognizes the telltale diagonal modulations
of tweed developing in the pretransitional regime.  The simulation
reveals that the system does indeed accomodate the energetic demands
of the compositional disorder by generating a deformation as shown,
i.e. the tweed modulation is the natural response of the system to the
disorder.

The configurations in Figure~2 are generated by simulating the
continuum system described above, discretized onto a 51 $\times$ 51
mesh.  The simulation variables are the displacements $U_x$ and $U_y$
at each site, and using a finite difference scheme the strains
and strain gradients are found for use in calculating the free energy
\(free). A random composition field which varies around some average
concentration $\overline{\eta}$ is assigned at the beginning and held
static.  The full rotationally invariant strain tensor is calculated
from any arbitrary displacement field, and then used to find the total
energy of the system.  A Monte Carlo simulated annealing algorithm is
used to minimize this energy, and generate a stable low energy
configuration for a given point in parameter space.  Typically, we
quench over four decades of temperature, using three thousand Monte
Carlo steps per lattice site per decade.


The materials parameters used in the simulation are those appropriate
for FePd.  Static harmonic elastic constant
measurements\refto{Muto_Softening} have given us $A_\phi\ (2.5 \cdot
10^{10}$ N/m$^2$ at the onset of tweed), $A_2\ (28 \cdot 10^{10}$
N/m$^2$), and $A_1\ (14 \cdot 10^{10}$ N/m$^2$). The strain gradient
parameter $\kappa/a^2\ (2.5
\cdot 10^{10}$ N/m$^2)$ can be calculated from the curvature of the
TA$_1$ phonon dispersion curve\refto{straingradcoeff}.  The
coefficients $\beta\ (1.7 10^{13}$ N/m$^2$) and $\gamma\ (3 \cdot
10^{16}$ N/m$^2$) are determined by the martensitic strain and the
value of $A_\phi$ at the transition.  The coupling to temperature,
$A_T\ (2.4 \cdot 10^8$ N/m$^2$ K), is known from
measurements\refto{Muto_Softening} of the temperature dependence of
$A_\phi$.  The coupling to statistical compositional variations,
$A_\eta$, will be discussed in detail below.

\com{
The mesh size of the simulated system can correspond to the lattice
scale, but does not have to.  Since computer power is at a premium,
and since we are investigating a mesoscale behavior, it is more
efficient to only descretize down to a mesh size which corresponds to
the length scale of the deformations.  Indeed, we must discretize at
least down to this scale in order for a continuum approximation to
remain valid.  The elastic constants $A_\phi, C_{44}, C_{12}, \beta$ and
$\gamma$ do not depend on the scale.  However, $\kappa$ must be
divided by the mesh size squared, to account for the scaling of the
gradient term, and $A_\phi_\eta \cdot \delta\eta$ must be divided by the
mesh size, to account for the scaling of the $1/\sqrt{N}$ statistical
composition variations.  We have chosen a value of the mesh size of
ten lattice constants, which allows computation of a sufficiently
large system yet maintains a valid continuum approximation.
}


A phase diagram, Figure~3, generated by the simulation, is
straightforward and intuitively sensible.  The vertical axis is the
elastic constant $\bar{A_\phi}$ at the nominal composition, i.e. the
``average'' value of the elastic constant.  Since $\bar{A_\phi}$
softens linearly with temperature over a large temperature range (at
least $150^\circ K$), this axis also effectively reflects temperature.
The horizontal axis\refto{phasediagram} is the strength of the
coupling, $A_\eta$, between the strain order parameter $\phi$ and the
composition inhomogeneity, $\delta\eta$.  The general structure of the
phase diagram is good confirmation of the general mechanism underlying
our model: sufficiently far from the thermodynamic transformation
temperature for the nominal compositon, the expected conventional
phases appear, while near to the transformation temperature there is a
region where the effect of the disorder becomes important, the lattice
deforms, and tweed appears.

In experimental observations\refto{Tweed_FePd}, as the temperature of
a sample is lowered toward the martensitic transformation temperature,
a smooth and unremarkable TEM image gives way to a mottled pattern
which signals the onset of some static lattice distortion --- static
at least on the time--scale of TEM observations.  With further
decrease in temperature, the mottled pattern organizes into a pattern
with a distinguishable directionality, acquiring a noticeable but
diffuse tweediness.  As the transformation temperature is approached,
the tweed develops increasingly coarse and long--range correlations,
and as the sample passes through the martensitic transformation, the tweed
gives way to fully transformed martensite, perhaps nucleating the
emerging finely twinned structure\refto{twinsform}.

In direct correspondence with these experimental observations, the
simulation yields precisely this same progression of pattern
development as $\bar{A_\phi}$ is decreased, (where $A_\eta$ is held
fixed at some constant value).  A perfectly undeformed system is
initially interrupted by scattered, non-interacting and uncorrelated
regions of distortion.  These are regions which have relatively large
values of $\delta \eta$ (large negative values, since $A_\eta$ is
positive for FePd) and are therefore the first to transform from
the (locally metastable or eventually unstable) austenite phase.
Within the constraints of the surrounding lattice, they thus deform
precociously toward the martensitic phase, giving the system a mottled
appearance. (This is the region above the austenite-tweed boundary.)
As $\bar{A_\phi}$ is further lowered, these regions grow dense enough
to interact, and longer range diagonal correlations develop, yielding
a diffuse tweed which grows increasingly distinct as $\bar{A_\phi}$ is
further lowered.  As $\bar{A_\phi}$ approaches the nominal
transformation temperature of the sample, an increasing fraction of
the sample prefers the martensitic phase, and the tweed grows very
coarse before finally transforming into the twinned martensite
configuration (which lies below the tweed--martensite boundary.)  The
precise placement of the austenite--tweed boundary is somewhat
ill-defined, as the distinction between ``correlated'' and
``uncorrelated'' is subjective, particularly in small samples such as
those studied here.  (A quantitative study of the degree of
correlation can be found elsewhere\refto{KarthaThesis}.)  The
tweed--martensite boundary, however, is well-defined, as the onset of
long--range martensitic order is a qualitative transition that can be
located precisely.

The parameters in this phase diagram are $\bar{A_\phi}$ (or
equivalently, temperature) and the strength of the coupling to
compositional variations, $A_\eta$.  In the laboratory, temperature is
easily varied, but for any given alloy the coupling strength is an
unadjustable property of the material, so the behavior of a sample
will trace a trajectory through the phase diagram which falls along a
single line, presumably with essentially constant $A_\eta$.
(Alternately, holding temperature fixed, the behavior of a material
may be investigated over a range of $\bar{A_\phi}$ by studying samples
of varying nominal compositions, $\bar{\eta}$.)  By comparing the
electron microscopy observations of FePd to the phase diagram derived
from simulations, we determine the effective value of $A_\eta$.  In
experimental investigations of FePd\refto{Muto_Softening} the onset of
tweed is seen to be roughly one hundred degrees above the
transformation temperature, corresponding to $A_\phi = 2.5 \cdot
10^{10} {N \over m^2}$.  By matching to the experimentally observed
tweed range, we determined the strength of the coupling to composition
variation required to generate tweed over this range in our
simulation; the value found is $A_\eta \approx 2.0 \cdot 10^{10} {N
\over m^2}$.  This figure may be compared to the following somewhat
simplistic estimate for $A_\eta$.  If, say in a binary alloy such as
FePd, the full ``bulk'' composition variation ($d \bar{A_\phi}/ d
\bar\eta$) coefficient were assigned to each lattice site, and the
composition at each simulation site varied between pure Fe or Pd, the
statistical fluctuation of $A_\phi$ would be 50 times larger than that
found to be required in our simulations.  This result should be
regarded in the light that there is apparently plenty of driving force
provided by simple compositional variations to produce tweed, even in
the absence of any specific defects or order--disorder changes.

\com{
This discrepancy in magnitudes draws attention to the almost overly
general nature of the Landau--Ginzburg approach.  The power of the
approach is its very global applicability, which allows us to neglect
system--specific aspects of the problem for the meantime, while
remaining confident that the important physical aspects are captured
in the model.  The problem of precisely determining the materials
parameters and other required coefficients may therefore be left until
later, when one attempts to extract exact quantitative information.
In a rigorous effort to determine a parameter such as $A_\eta$, the
coupling to compositional variations, several important issues beyond
what we have discussed will have to be addressed:}

This rather large estimate for the coupling $A_\eta$ neglects
several important points which should be considered in any careful
attempt to calculate the coupling to composition: 1) $d \bar{A_\phi}/ d
\bar\eta$ is the product of $d T_M / d \bar\eta$ and $d A_\phi/ d
T$, where each of these are known from experimental measurements near
$T_M$.  Linearly extrapolating away from the range of $\bar\eta$ (29\%
to 32\%) over which the martensitic transformation occurs may well
overestimate its strength for concentrations outside of this range.
As mentioned above, it is convenient to think of the martensitic
transformation as occurring at a (weakly temperature dependent)
critical composition, so it is perhaps more accurate to consider
$A_\phi(\bar\eta)$ as a step function at the critical composition, say,
rather than a simple linear function.  2) As will be discussed further
below, finite temperature effects are important in these systems.
Thermal lattice vibrations are quite substantial at the temperatures
at which tweed is seen, and will be correlated over some temperature
dependent length scale, transmitting and averaging out the effects of
any disorder, including local compositional variation.  3) In any
Landau--Ginzburg theory of a non-uniform system, the existence of a
local free energy, (a concept which is thermodynamic in nature,)
implicitly assumes that one has ``integrated out'' certain (secondary)
degrees of freedom.  For example, defining a free energy functional of
a static strain tensor requires integrating over phonon modes, which
necessarily introduces a coarse--graining length
scale\refto{KrumhanslGooding, BruceCowley}.  Introducing this length
scale into the problem will result in averaging the effect of
compositional variations (and will also change other parameters of the
model).  4) Any physical mechanism which is based on the chemistry of
an alloy or doped compound will involve electronic effects which will
exert their influence over a length scale which is larger than the
lattice spacing, typically on the order of a Fermi length.  Again,
this will lead to a spatial averaging of composition, weakening the
apparent strength of the coupling to local compositional variation.
5) This two dimensional model, although faithful to the real three
dimensional material from the perspective of symmetry requirements,
neglects an important effect of dimension on compositional
fluctuation: a composition field which is defined by averaging within
some radius will average over a region of material which whose size
will depend on dimension.  Correspondingly, in a higher dimension
there will be smaller compositional fluctuations.

\com{
({\it jimk's comments}) is undoubtedly due to the omission of the
intrinsic coarse--graining in the free energy which must average
variations over a number of lattice sites, thus reducing the
statistical variance of the elastic free energy $A_\phi$.  The microscopic
determination of the coarse--graining scale is a subject for further
investigation.  For now, our model can be taken as an internally
self-consistent way to deduce (by simulation) the manner in which the
$\phi$ elastic free energy varies in space with compositional
variation.
}

\com{
({\it remarks above replace:} This value may be compared to an upper
limit on the driving force that is hypothetically available due to
compositional variations, based on assuming a simple relationship to
the known composition dependence of the transformation temperature,
$A_\eta \equiv {d A_\phi \over d T} \cdot {d T_M \over d \overline\eta}
\approx 10^{12} {N \over m^2}$.  Since this is larger by a factor of
50 than the required coupling, we can be reassured that there is
indeed sufficient driving force provided by compositional variations
to generate tweed.
}
\noindent{\bf Spin Glass and Tweed}\hfil\break
The dotted lines in the phase diagram (Figure~3) are drawn to provide
a comparison between the model presented here and earlier
work\refto{Kartha} in which the martensitic system was treated
analytically by taking the approximation of infinite elastic
anisotropy\refto{Ericksen} and formally mapping tweed onto a spin
glass system.  The infinite anisotropy approximation is motivated by
the observation of severe softness of $A_\phi$ in many martensitic
materials, and a corresponding growth of the elastic anisotropy,
$\alpha \equiv C_{44} / C'$.  (For example, in FePd $\alpha \sim 20$,
in NiAl $\alpha \sim 10$, and in some Indium alloys $\alpha$
approaches $\sim 300$ !\refto{Gobin}.)  As discussed fully below,
solutions in this approximation are given by displacement fields of
the form $${\bf U}(x,y) = \pmatrix{ 1\cr -1\cr} U_+(x+y)+ \pmatrix{
1\cr 1\cr} U_-(x -y)\eqno(constraineddisplacement) $$ where $U_+$ and
$U_-$ are arbitrary functions of position along the $\langle 11
\rangle$ and $\langle 1\bar{1}\rangle$ directions.  These limiting
solutions clearly have infinitely long correlations in the diagonal
directions, explaining why there is natural tendency for a tweed-like,
diagonal modulation.  (This tendency for a displacement field is
related to the fact that twin boundaries appear only along $\langle 1
1 \rangle$ directions.)  Although this is clearly a very severe
constraint on the form of the allowed solutions, such displacement
fields are surprisingly still capable of smoothly and continuously
modulating between regions of the high and low temperature phases.
That is to say, regions of the austenite and the two martensitic
variants can be patched together without generating any of the two
energetically costly strains, $e_1$ and $e_2$.  As in the full elastic
model of this paper, it is compositional inhomogeneities which
generate a local propensity toward one phase or the other, and thereby
provide the driving force behind the tweedy modulation.

The approximation of infinite elastic anisotropy helps to explain how
the lattice manages to accomodate the compositional disorder, but
further, it demonstrates the subtlety of that accomodation.  In the
effort to adjust to the local disorder, subject to the constraint of
infinite diagonal correlations, the displacement field suffers
substantial frustration.  As in many systems, this coupling of
disorder (compositional) and frustration (elastic strain) gives rise
to glassy behavior.  In a formal and rigorous way, this claim can be
made mathematically precise, and the martensitic system can be mapped
to an infinite range bipartite Sherrington--Kirkpatrick spin glass.
(See \ref{Kartha} for details.)  Tweed is therefore an intermediate
phase between the high temperature square phase and the low
temperature rectangular phase, in direct correspondance with the spin
glass phase which exists between the ferromagnetic
and antiferromagnetic phases.

Admittedly, actual materials do not have {\it infinite} elastic
anisotropy: one should therefore not too boldly assert claims founded
on what amounts here to a mean--field approximation.  Tweed, which in
this approximation is an actual thermodynamic phase with second order
phase boundaries, will appear in the real world as a ``ghost'' of an
intermediate phase, perhaps without true long-range order in time, but
with observable glassy behavior.  It is therefore of particular
interest to investigate the nature of the remaining glassy behavior
when we relax the approximation of infinite elastic anisotropy.

	The numerical simulation, which uses materials parameters
appropriate for FePd, allows us to gain insight into the nature of the
glassy manifestations in a system with realistic finite elastic
anisotropy.  In particular, we can observe the transition to glassy
behavior in our numerical simulation as temperature is decreased.
Note, as explained above, temperature is introduced into the
simulation through the temperature dependence of the softening elastic
constant $A_\phi$.  However, the Monte Carlo method we have adopted
allows us to also introduce thermal fluctuations and study their
ability (or inability, as the case may be) to destroy the long range
order in time which is the signature of a glassy system.  Using the
number of attempted Monte Carlo steps per site (MCS) as a measure of
time, we have identifued a regime in phase space where fluctuations
become very slow.

	To quantify such glassy behavior, we have measured time
correlations of the martensitic $\phi$ distortion. Because this strain
fluctuates around a zero mean, the quantity\refto{EA}: $$\xi(t)
\equiv {1 \over N} \sum_i\> \phi(site\; i,\; time\; 0)\,
\phi(site\; i,\; time\; t)$$ will be zero unless the values of
$\phi$ at time zero and time t are correlated. Figure~4 shows values
for this correlation function, after normalizing to $\xi(0) = 1$ and
averaging over a number of intervals for each time t for improved
statistics. All of the data in the graph were produced with a single
set of parameters which gave clear tweed when thermal fluctuations
were negligible. The physical temperature given for each curve is
determined from the Monte Carlo temperature by using the known elastic
constants, strains, and the appropriate grid spacing to equate the
Monte Carlo fluctuations with a thermal Boltzmann distribution.


	The behavior at the two temperatures differ sharply. At 38 K,
memory of the deformations present at t=0 lasts over tens of thousands
of steps, as some local distortions have been unable to surmount the
potential barrier between one martensitic variant and the other.  At
380 K, however, all areas in the lattice are able to fluctuate between
both martensitic variants, and memory of the original configuration is
nearly lost after merely a hundred MCS. Using a typical frequency for
atomic motion and the size of each Monte Carlo step (which was held
constant in this part of the study), we can estimate that a thousand
Monte Carlo steps corresponds to several picoseconds.  Since the time
scale for neutron scattering is of order picoseconds, we emphasize
that there exists a regime in the numerical simulations for which the
resulting deformations constitute a robust tweed pattern which is
static for a duration which is experimentally significant.

	Clearly, the low temperature model in Figure~4 has a wide
distribution of relaxation times, characteristic of glassy behavior.
If our system were a true spin glass on an infinite lattice, the
longest relaxation time scales would diverge and there would be true
long range order in time\refto{spinsbeyond}. Experimental observations
of physical systems which display spin glass behavior have observed
relaxation times of days or weeks\refto{spinsbeyond}. Even spin glass
simulations on small lattices have relaxation spectra which are
bounded above only by the (rather large) time scale associated with
flipping all of the spins in the lattice.  To get a grasp of how long
such waiting times might be in our system, recall that the mapping
from a spin glass to our infinite anisotropy model maps one spin to a
correlated region as long as a lattice diagonal, so that the energy to
flip over a ``cluster" of our ``spins" is really the overwhelming
elastic energy barrier to reversing strain in several long,
overlapping tweed strips.  In our finite anisotropy limit, correlated
tweed regions are still large compared to the lattice constant, so the
dynamics in our model will still resemble that expected in a (large
but finite) spin glass\refto{myexp}. It seems likely that the
hysteresis\refto{Oshima_Hysteresis} and frequency dependent
relaxation\refto{frequency_dependence} seen in the tweed regime in
real materials hints at the glassy, slow dynamics predicted by our
model.

\noindent{\bf Diffraction}\hfil\break
Transmission electron microscopy studies are not always designed to
yield direct information about the underlying atomic displacements.
It is therefore useful to refer to the diffraction images produced in
conjunction with the tweed observations.  Figure~5a shows
experimental\refto{Moss} x-ray diffraction contours taken from tweed
produced in YBaCuO, and Figure~5b shows diffraction patterns at the
same Bragg peaks for the simulated tweed computed directly by Fourier
transforming the displacements in the simulation.  As can be seen, the
qualitative features are faithfully reproduced.


	Beyond the simple reassurance derived from qualitatively
matching simulated diffraction patterns with experimental results,
additional information comes from focusing on reciprocal space since
real space strain images and reciprocal space diffraction patterns
yield different information under the time averaging present in any
data collection process. In a real space image which makes one
martensitic variant light and another dark, a region which is
fluctuating quickly from one variant to the other would appear to be
grey in a coherent time average.  (This could be the case for a
tetragonal to orthorhombic transformation observed under ``two-beam"
TEM imaging, which shows strain projected along a given direction.)
In a diffraction pattern from sample of tweed, the characteristic
cross pattern is a result of {\it correlations} in strain, not just
the {\it absolute} strain.  Even if the specific martensitic
deformations vary in time, the incoherent time average which is given
in a diffraction pattern will still show clear streaks as long as the
strains maintain persistent instantaneous correlations.


	Figure~6 demonstrates that ``static'' tweed (upper row) melts
to a ``dynamic'' tweed (lower row) as temperature is
increased\refto{datafrom}.  The first column shows the instantaneous
configuration: the distinctly tweedy deformation in the upper sample
is largely obscured in the lower sample by thermal fluctuations.  The
middle column shows an average of real-space deformations over 150,000
attempted Monte Carlo steps (corresponding to averaging over
approximately a nanosecond).  The tweed in the upper sample is still
easily discerned, whereas the lower sample has averaged to virtually
zero net deformation.  However, both simulated samples yield a
time--averaged diffraction pattern with the characteristic diagonal
streaking, revealing the presence of instantaneous tweedy
correlations.  This is analogous to the presentation of Van Tendeloo
{\it et al.}\refto{LeSar}, who show that the tweedy correlations of
static and dynamic tweed can be verified by their diffraction images,
even though they behave very differently over the relatively long time
required to make a TEM image.

	Of course, ``static" and ``dynamic" are only defined on the
time scale of the relevant real space observations. However, our
theory predicts a clear transition to macroscopically long range order
in time, and current evidence supports this prediction.  Molecular
dynamics simulations of tweed in NiAl have found a change from static
to dynamic behavior with increased temperature\refto{Becquart}, and
neutron diffraction and TEM observations of real NiAl find a static
component of tweed which appears and then grows as the temperature is
lowered\refto{TannerStatic}. In our own numerical simulations, we've
observed such a transition by detecting the onset of correlations
below some temperature.  For example, the static tweed of Figure~6 is
static (on a time scale of several thousands MCS) from 0K to roughly
70K, and dynamic from 70 K to 90 K.  Similarly, the dynamic tweed of
Figure~6 actually arises from a sample which is fully twinned up to
roughly 76K, and then is a dimly visible (i.e.
static\refto{hardnumber}) tweed at 85K, and then is dynamic at
temperatures as high as 114K.  Quantitatively, this same transition is
demonstrated in Figure~4.

Thus, in our model, both static and dynamic tweed are present over
temperature ranges of tens of degrees Kelvin. In a real three
dimensional sample, one can assume there would be an even larger tweed
range, since thermal fluctuations are more heavily damped in three
dimensions.  The simulation results also serve as a reminder that,
depending on the material in question, either ``static" or ``dynamic"
behavior may dominate most of the temperature range of tweed. Since
available experimental probes react differently to changes in
dynamical behavior, understanding variations in the time dependence of
tweed will be vital for appropriate comparison of data.  These
concerns apply not only to diffraction patterns and TEM images, but
also to the ``central peak" of inelastic neutron scattering, which has
been associated with tweed which is static on the time scale of
neutron scattering\refto{TannerStatic}.  Interestingly, our findings
also indicate that introducing thermal fluctuations can produce a
transition from martensite to tweed, once again implicating the long
suspected influence of vibrational entropy as a stabilizing effect for
the high temperature phase.

It has long been known\refto{Tweed_CuBe} that the diffraction patterns
such as those shown above are consistent with the presence of
$\{110\}$ planes shearing in $\langle 1\bar 10 \rangle$ directions,
the so--called Zener mode\refto{Zener}.  The square $\rightarrow$
rectangular transformation considered here, as well as the cubic
$\rightarrow$ tetragonal and tetragonal $\rightarrow$ orthorhombic
transformations all result from precisely such $\{110\}/\langle
1\overline 1 0\rangle$ shears.  In addition, this shear (coupled with
an additional homogeneous strain) is responsible for body--centered
cubic $\rightarrow$ close--packed transformations.  The observation of
a pretransitional deformation which involves this particular shear is
therefore very consistent with the approach of the martensitic
transformation.  Furthermore, it is this shear mode that couples to
the elastic constant $A_\phi$ which is seen to soften in many
materials as the martensitic transformation temperature is approached
and which motivated the approximation of infinite elastic anisotropy
discussed above.

For completeness, let us make explicit the connection between the
observed diffraction behavior and the lattice displacements.  To
consider the diffraction pattern from a distorted lattice, we make use
of the fact that an arbitrary displacement field can be written in a
perfectly general way as $${\bf U}(x,y)=\pmatrix{1\cr -1\cr}
U_+\left({{x+y}\over d},{{x-y}\over L}\right)+
\pmatrix{ 1\cr 1\cr} U_-\left({{x+y} \over L}, {{x-y} \over d}\right).
\eqno(generaldisplacement)$$
In the infinite anisotropy approximation, $L$ would be taken to be
infinity, yielding eq.\(constraineddisplacement).  A displacement
consisting of long (but not infinite) diagonal correlations can be
expressed by taking $L \gg d$ (where we define $U_-$ and $U_+$
such that they have similar functional dependences on their first and second
arguments.)  The correlation length $L$ then describes the
longitudinal length scale of the tweed striations, and $d$ describes
their transverse width.

The general expression\refto{Krivoglaz} for the diffracted intensity
at a wavevector ${\bf Q}$, (scattering from sites ${\bf s}$ with
structure factors $f_{\bf s}$) is $$ I({\bf Q}) = \vert \sum_s
f_s({\bf Q}) e^{i{\bf Q}\cdot {\bf R}_s}\vert^2. \eqno(intensity)$$ If
we write ${\bf Q} = {\bf K} + {\bf q}$ (where ${\bf K}$ is the nearest
reciprocal lattice vector to ${\bf Q}$) and write ${\bf R}_s = {{\bf
R}^\circ}_s + {\bf U}_s$ (where ${\bf U}_s$ is a displacement around
the lattice point ${{\bf R}^\circ}_s$) then we can expand \(intensity)
assuming small displacements and find: $$I({\bf Q}) =\vert f({\bf Q})
\vert^2 \vert
\sum_s ( {\bf Q}\cdot{\bf U}_s) e^{i {\bf q}\cdot {{\bf R}^\circ}_s}
\vert^2$$
(where for convenience we have assumed a monotomic lattice, or at
least a lattice with an effectively constant $f_s$). Fourier expanding
${\bf U}_s$ and noting that the summation over sites will give us a
delta function, we find: $$ I({\bf Q})=\vert f({\bf Q})\vert^2
\vert{\bf Q}\cdot{\bf U}_{\bf q}\vert^2.\eqno(intensityapproximate)$$

Eq.\(intensityapproximate) indicates that, in the approximation of
small displacements, the intensity of diffuse scattering around Bragg
peaks (normalized by $1/\vert f({\bf Q})\vert^2$ will obey a $|{\bf
Q}|^2$ dependence.  Comparing the observed scattering intensity to a
$|{\bf Q}|^2$ fit will allow us to verify that the scattering is
caused by a {\it small} lattice deformation.  Substantial deviation
from a strict $|{\bf Q}|^2$ dependence would imply that the
approximation of scattering from small displacements is not
appropriate, suggesting that the scattering is due perhaps to
substitutional disorder or to microdomains large enough to produce
size broadening\refto{Moss}.  Figure~7 shows the diffuse scattering
intensity (normalized by the appropriate structure factor) plotted for
Bragg peaks $\langle 0\ Q\ 0\rangle$, with $Q / (2 \pi /a) = 4,6,8$,
and 10 for the experimental measurements\refto{Moss} (filled circles)
and $Q / (2 \pi /a) = 0,1,2,...10$ for our simulated diffraction
patterns (open circles).  The curve shows the best $|{\bf Q}|^2$ fit
to the experimental data.  The minor deviation from the $|{\bf Q}|^2$
fit reflects the fact that displacements are finite, yet small.


Assured that we are indeed observing diffuse scattering from small
displacements, we can proceed by solving the Fourier transform of
eq.\(generaldisplacement), $\tilde {\bf U}(\bf k) = \tilde {\bf
U}_+(\bf k) + \tilde {\bf U}_-(\bf k)$, where $$\tilde {\bf U}_+({\bf
k}) = \int {dx dy \over \sqrt A}
\pmatrix{1\cr -1\cr} U_+({{x+y}\over d},{{x-y}\over L})
\ e^{i\ {\bf k} \cdot (\bf x + \bf y)}\eqno(Fourier a)$$
$$\tilde {\bf U}_-({\bf k}) = \int {dx dy \over \sqrt A}
\pmatrix{1\cr 1\cr} U_-({{x+y}\over L},{{x-y}\over d})
\ e^{i\ \bf k  \cdot (\bf x + \bf y)}\eqno(Fourier b)$$
Make the substitutions $k_\pm \equiv (k_x \pm k_y) / \sqrt 2$ and $$ s
\equiv {(x + y) \over d \sqrt 2},
\qquad\hbox{and}\qquad t \equiv {(x - y) \over L \sqrt 2}$$
in eq. \(Fourier a) and $$ s \equiv {(x + y) \over L \sqrt 2},
\qquad\hbox{and}\qquad t \equiv {(x - y) \over d \sqrt 2}$$
in eq. \(Fourier b).  We then find $$\tilde {\bf U}_+({\bf k}) = {Ld
\over \sqrt A} \int ds dt
\pmatrix{1\cr -1\cr} U_+(s,t)
\ e^{i\ (d k_+ s\ +\ L k_- t)} \eqno(Fourierswitch a)$$
$$\tilde {\bf U}_-({\bf k}) = {Ld \over \sqrt A} \int ds dt
\pmatrix{1\cr 1\cr} U_-(s,t)
\ e^{i\ (L k_+ s\ +\ d k_- t)}\eqno(Fourierswitch b)$$
The general shape of this Fourier transform is clear by inspection:
expression \(Fourierswitch a) is simply a Fourier transform of
$U_+(s,t)$ {\it scaled} by $1/d$ along the $\hat {\bf s}$ direction
and $1/L$ along the $\hat {\bf t}$ direction (similarly for $U_-$).
Since, by construction, $U_+$ and $U_-$ are simple isotropic
displacement fields, then so are their Fourier transforms, and the
final result is a Fourier transform, $\tilde {\bf U}(\bf k)$ which has
the shape of two diagonal streaks emanating from the origin, the
length of each streak being $\sim 1/d$, and the width $\sim 1/L$.
In addition to the variation in diffuse scattering with $|{\bf Q}|$
discussed earlier, there is also a marked variation with the
orientation of ${\bf Q}$, as demonstrated in Figure~5. The product
${\bf Q} \cdot{\bf U}_{\bf q}$ in Eq.\(intensityapproximate) will
cause diffuse scattering resulting from $U_+$ to vanish for ${\bf Q}
\parallel (1,1)$ and the diffuse scattering from $U_-$ to vanish for
${\bf Q} \parallel (1,-1)$.  This is known as the ``extinction
condition'' associated with $\{110\}/\langle 1\overline 1 0\rangle$
shears.  (\Ref{KarthaThesis} discusses extracting $d$ and $L$ from data.)

Figure~8 shows the trace of the diffuse scattering around the Bragg
peak (660) in the $\langle \bar 1 1 0 \rangle$ direction for
YBaCu(Al)O.  The corresponding simulation diffraction data is also
shown, averaged over several simulation runs for improved statistics
and scaled by a single constant for both the $\langle \bar 1 1 0
\rangle$ direction and the $\langle 1 1 0 \rangle$ direction.  The
simulation data points closely follow experimental points for the
regime covering the first two orders of magnitude of intensity, but
then fall off quickly, partly due to the absence in the simulated
diffraction pattern of background scattering which is experimentally
unavoidable.  As Jiang {\it et al.}  have shown\refto{Moss} for
YBaCu(Al)O, the $1/q^2$ behavior indicated by the line corresponds to
a lattice distortion which can be explained as a linear elastic
response in a material with a random distribution of oxygen atoms.
The fact that the intensity falls off quicker than $1/q^2$ is offered
by them as evidence for non-random short range ($\sim 40$ \AA)
correlation in the oxygen ordering.  We attach no great significance
to the agreement between experiment and our simulations: our theory is
non-linear, but far from the transformed regions (small $q$) the
strain field is dominated by linear response.  It is simply reassuring
that the computed structure of the scattering profile is consistent
with experiment.


\bigskip
\noindent{\bf Compatibility and Nonlocal Interactions}\hfil\break
Having reviewed the explicit relationship between the tweed deformation
and the observed diffraction pattern, we wish now to explain the physics
underlying that deformation.  We wish to repeatedly emphasize that an
essential aspect of the physics underlying tweed is that any local
fluctuation in the elastic free energy functional (e.g., due to
compositional inhomogeneity) cannot be regarded in isolation.  The
lattice response is not simply a superposition of the responses
expected for independently considered sites of disorder.  Rather,
mutual non-local interactions between spatially separated regions
conspire to give an extended cooperative response: i.e. the tweed
phenomenon.  In this section, we here develop a simple yet
illuminating analysis which explicitly uncovers the effective
non-local interaction which leads to these extended cooperative
responses.  We write the non-local interaction in terms of a
renormalized Fourier space elastic constant, and show that the
specific form of the tweed deformation immediately follows.

The only explicitly non-local term in the free energy Eq. \(free)
is the strain gradient term $(\nabla \phi)^2$.  However, even
disregarding this term, the order parameter $\phi$ cannot be an
entirely arbitrary function of position: there is no guarantee that
such a field (even if continuous) is physical.  This problem relates
to the following important subtlety of any Landau-Ginzburg-type model
which treats elastic strain as the relevant order parameter: the true
degrees of freedom in a continuum elastic medium are contained in the
{\it displacement} field, ${\bf U}({\bf x})$, even though it is the
{\it strain} fields, $e_{ij}$, which appear in the free energy.
Instead of treating the strains as independent fields, one must assure
that they correspond to a physical displacement field, i.e. that they
are derivatives of a single continuous function.  This is done by
requiring that they satisfy a set of non-trivial {\it compatibility
relations}\refto{Sokolnikoff} concisely expressed by the
equation\refto{Baus} $\nabla \times (\nabla \times \bf e)^\dagger =
0$.  In two dimensions this can be written as
$$\nabla^2 e_1 - \sqrt 8\ \partial_{xy}\ e_2 - \Bigl(\partial_{xx} -
\partial_{yy}\Bigl) \phi = 0.\eqno(compatibility)$$

Ignoring this geometrical compatibility constraint and minimizing the
free energy directly would lead to the incorrect result that $e_1$ and
$e_2$ are identically zero, and $\phi$ (the only field directly
coupled to the composition) trivially responds to the local disorder.
We can explicitly account for the compatibility constraint by
appending it to the free energy \(free) via a Lagrange multiplier,
$\lambda({\bf x})$.  It is then possible to solve for $e_1$ and $e_2$
because we have two constraints relating the three strain fields: the
compatibility condition \(compatibility) and the requirement that the
free energy is minimized.  Solving for $e_1$ and $e_2$ in terms of the
order parameter $\phi$, we are able to express the free energy in
terms of $\phi$ alone.  The contributions of $e_1$ and $e_2$ to the
free energy will be accounted for\refto{guaranteed} by terms which
have the appearance of a non-local interaction coupling $\phi({\bf
x})$ and $\phi({\bf x'})$. Note, by integrating out $e_1$ and $e_2$
from the free energy in this way, we are not resorting to an
approximation of infinite anisotropy, we are simply analytically
solving for the fields $e_1$ and $e_2$ in terms of an arbitrary $\phi$
field.

Proceeding, we find the solutions of the two secondary strain fields
$e_1$ and $e_2$ which minimize the free energy for a given field
$\phi$ subject to the compatibility constraint.  In the standard way,
we extremize the free energy with respect to variations in the strain
fields, and find Euler-Lagrange ``equations of motion'' relating the
two secondary strain fields to $\phi$.  Introducing variations $\delta
e_1$, $\delta e_2$, and $\delta \lambda$ into \(free) yields
$$\delta f = A_1\, e_1\, \delta e_1 + A_2\, e_2\, \delta e_2 +
\lambda \Biggl\{\nabla^2 \,\delta e_1 -  \sqrt 8\ \partial_{xy}
\,\delta e_2 \Biggr\}$$
$$+ \delta \lambda \Biggl\{\nabla^2 e_1 -  \sqrt 8\ \partial_{xy}
\  e_2 - \Bigl(\partial_{xx} - \partial_{yy} \Bigr)
\phi\Biggr\}\eqno(Fvariation)$$
Doing the requisite integrations by parts, and requiring that $\delta
f$ is zero, we find the following Euler--Lagrange ``equations of
motion'': $$e_1 = {-1 \over A_1} (\nabla^2 \lambda)\eqno(eofm a)$$
$$e_2 = {\sqrt 8 \over A_2} \Bigl(\partial_{xy}\ \lambda
\Bigr)\eqno(eofm b)$$
where $\lambda$ is given by: $$ - {1 \over A_1} (\nabla^2)(\nabla^2
\lambda) -  {8 \over A_2} \Bigl(\partial_{xxyy} \ \lambda \Bigr) =
 \Bigl(\partial_{xx} - \partial_{yy} \Bigl) \phi\eqno(eofm c)$$

This opaque set of equations becomes quite transparent after
reexpressing in k-space: $$\tilde{\lambda}({\bf k}) = {(k_x^2 - k_y^2)
\over (k_x^2 + k_y^2)^2/A_1 + 8 (k_x^2 k_y^2)/A_1} \ \tilde{\phi}({\bf
k}) \eqno(kspace a)$$

$$\tilde{e_1}({\bf k}) = {(k_x^2 + k_y^2) (k_x^2 - k_y^2)/A_1 \over
(k_x^2 + k_y^2)^2/A_1 + 8 (k_x^2 k_y^2)/A_2}\ \tilde{\phi}({\bf
k})\eqno(kspace b)$$

$$\tilde{e_2}({\bf k}) = {-\sqrt 8 (k_x k_y) (k_x^2 - k_y^2)/A_2 \over
(k_x^2 + k_y^2)^2/A_1 + 8 (k_x^2 k_y^2)/A_2}\ \tilde{\phi}({\bf
k})\eqno(kspace c)$$

The free energy, which is now a functional of $\phi$ alone, now may be
expressed $$F =
\int\!\!\int d{\bf r}\ d{\bf r'}\ \  f_1({\bf r},
{\bf r'})\ \phi({\bf r})\ \phi({\bf r'}) +
\int\!\!\int d{\bf r}\ d{\bf r'}\ \  f_2({\bf r},
{\bf r'})\ \phi({\bf r})\ \phi({\bf r'})$$ $$+ \int
f_{local}\bigl(\phi({\bf r})\bigr) d{\bf r}\eqno(freenonlocal)$$ where
$f_{local}$ is the part of the free energy \(free) which is
explicitly dependent on $\phi$ alone.  The non-local interactions
$f_1$ and $f_2$ account for the free energy contribution due to the
$e_1$ and $e_2$ strain fields, respectively.  They are given by $$f_1
({\bf r},{\bf r'})
\equiv \int {d{\bf k} \over a} {A_1 \over 2} \Bigl[{ (k_x^2 + k_y^2)
(k_x^2 - k_y^2)/A_1 \over 8 (k_x^2 k_y^2)/A_2 + (k_x^2 + k_y^2)^2/A_1
} \Bigr]^2 e^{i {\bf k} \cdot ({\bf r} - {\bf r'})}
\eqno(nonlocalint a)$$ and
$$f_2 ({\bf r},{\bf r'})
\equiv \int {d{\bf k} \over a} {A_2 \over 2} \Bigl[{ \sqrt{8} (k_x k_y)
(k_x^2 - k_y^2)/A_2 \over 8 (k_x^2 k_y^2)/A_2 + (k_x^2 + k_y^2)^2/A_1
} \Bigr]^2 e^{i {\bf k} \cdot ({\bf r} - {\bf r'})}
\eqno(nonlocalint b)$$
where $a$ is the system area.  More transparently, we can write $$F_1
= \int {d{\bf k} \over a} {A_1 \over 2} \Bigl[{ (k_x^2 + k_y^2) (k_x^2
- k_y^2)/A_1 \over 8 (k_x^2 k_y^2)/A_2 + (k_x^2 + k_y^2)^2/A_1 }
\Bigr]^2 |\phi({\bf k})|^2
\eqno(nonlocalint a)$$ and
$$F_2 = \int {d{\bf k} \over a} {A_2 \over 2} \Bigl[{ \sqrt{8} (k_x
k_y) (k_x^2 - k_y^2)/A_2 \over 8 (k_x^2 k_y^2)/A_2 + (k_x^2 +
k_y^2)^2/A_1 } \Bigr]^2 |\phi({\bf k})|^2
\eqno(nonlocalint b)$$
These terms are thus simple harmonic terms, where the bare elastic
constants $A_1$ and $A_2$ are now ${\bf k}$--dependent.  The harmonic
term in $\phi$ is now
$$\int {d{\bf k} \over a} {A_\phi({\bf k}) \over 2} |\phi({\bf
k})|^2\eqno(harmonicfourier)$$
with the restoring force $A_\phi({\bf k})$ given by
$$A_\phi({\bf k}) = A_\phi + A_1\, Q_1({\bf k})^2 + A_2\, Q_2({\bf
k})^2\eqno(renormalizedc)$$
where $$Q_1({\bf k})\equiv { (k_x^2 + k_y^2) (k_x^2 - k_y^2)/A_1 \over
(k_x^2 + k_y^2)^2/A_1 + 8(k_x^2 k_y^2)/A_2 }\eqno(kdep a)$$
and $$Q_2({\bf k}) \equiv {\sqrt{8} (k_x k_y) (k_x^2 - k_y^2)/A_2 \over
(k_x^2 + k_y^2)^2/A_1 + 8 (k_x^2 k_y^2)/A_2 }.\eqno(kdep b)$$
The key feature of the harmonic free energy, Eq. \(harmonicfourier),
is the factor $(k_x^2 - k_y^2)$ burried in Eqs. \(kdep), which leads in
a natural way to a tweedy deformation, as follows.


One can immediately see from eq. \(kspace b) and Eq. \(kspace c) that
strains $e_1$ and $e_2$ will be generated by non-zero $\phi({\bf k})$,
{\it except} for Fourier components for which $k_x^2 - k_y^2 = 0$.
Equivalently, one can see from \(nonlocalint a) and \(nonlocalint b)
that only Fourier components of $\phi({\bf k})$ with $k_x^2 - k_y^2 =
0$ will incur no free energy cost through the terms $f_1$ and $f_2$,
and that their contributions to the harmonic restoring force
\(renormalizedc) go to zero.  As a result, even for finite elastic
anisotropy, non-diagonal Fourier components of $\phi$ with $k_x^2 -
k_y^2 \ne 0$ will be suppressed relative to diagonal components, which
will be increasingly prominent for increasing anisotropy.  The
resulting deformation is a tweedy modulation with long but finite
diagonal correlations.


The significance of this nonlocal interaction lies partly in the
anisotropy of the interaction, as explained above, but also in the
range of the interaction.  In Figure~10, we plot the numerically
integrated function $(f_1 + f_2)$ along the axial and diagonal
directions.  The interaction strength along the axial directions is
positive, strongly suppressing axial correlations.  The interaction
along the diagonal direction is negative, enhancing diagonal
correlations but only insofar as this interaction survives the $k_x^2
- k_y^2$ suppression.  (Note, we plot in Figure~10 the absolute magnitude
of the interaction strength along the diagonal direction for better
comparison).  As one can see from the figure, this long--range
interaction dies off like $1/r^2$, and would produce logarithmic
divergences in a poorly accomodated two dimensional system.  The
inevitable result is the development of extended, coordinated lattice
modulations which take advantage of the compositional disorder (or any
other driving force which couples to the martensitc strain) without
generating unnecessary strains: i.e. tweed.

\bigskip
\noindent{\bf Alternative Couplings}\hfil\break
In the preceeding section, we have reexpressed the simple physics
embodied in continuum elasticity in a manner which emphasizes the
long--range, collective nature of the lattice's response to a
perturbing force.  In particular, this approach helps to clarify why a
tweedy modulation is the natural response of a system in which a
disorder field is coupled to the strain.  We will here briefly
consider possible couplings between the disorder field and the strain,
and argue that the coupling incorporated into the present model, i.e.
the term $\eta\phi^2$ is most effective at generating tweed and
most appropriately describes the experimental observations.

We have noted that the simplest coupling between the martensitic
strain $\phi$ and the composition $\eta$ is $\eta \phi^2$, prompting
its inclusion in our Landau free energy via the term $A_\eta\ \delta
\eta\ \phi^2$.  While it is true that this is the simplest coupling
between a scalar field and the order parameter $\phi$, our decision to
consider a {\it scalar} field coupled directly to the $\phi$ component
of the strain field requires justification, since non--scalar disorder
fields and the other strain components can not in principle be
neglected out of hand.

First, we've chosen to couple to $\phi$ directly for the same reason
that we've taken it as our order parameter and included its anharmonic
terms: $\phi$ is the predominant strain measured in the tweed
deformation and it is the strain responsible for the martensitic
transformation.  As such, it is larger in magnitude than the other
strain components and will be most susceptible to interacting with a
driving force, whether due to intrinsic disorder or some extrinsic
field such as an externally applied stress.

Second, we've chosen a scalar disorder field simply because tweed is
so impressively widespread a phenomena, and we desire to study the
simplest, most universal disorder.  Indeed, there are important
materials with unit cell configurations that allow for some disorder
field more complicated than simply scalar: the obvious example is
oxygen concentration in YBCO.  However, we find it provocative that
tweed appears in materials with even simple lattices such as FCC or
BCC.  Here, disorder takes the form of random placement of atoms, with
each site being symmetrically equivalent with every other.  In this
case, disorder is necessarily simply a scalar field for which the
broken symmetry of the martensitic strain precludes a linear coupling.
Even in the case of nickel--rich Ni$_x$Al$_{1-x}$, which has an
ordered $\beta$--CsCl structure, the compositional disorder can be
described by a scalar field corresponding to the positions of the
excess nickel atoms, which are accomodated by random substitution onto
the aluminim sublattice\refto{Bradley}.

Third, we know that composition couples not only to the martensitic
transformation strain, but to the martensitic transformation {\it
temperature}.  Compositional inhomogeneities will result in a
spatially varying transformation temperature, and in a $\phi^6$ Landau
free energy this is reflected in spatial variations in the coefficient
of $\phi^2$.

Fourth, the experimental observation of hysteresis immediately allows
us to conclude that a simple linear response mechanism cannot be the
general origin of tweed.  Hysteresisis is
seen\refto{Oshima_Hysteresis} to occur upon cycling of temperature:
upon heating, the tweed pattern persists up to a temperature which is
higher than that at which it initially appeared upon cooling.  This
hysteresis implies that the tweed is something more complex than
simply linear response of the lattice to some static defect or
impurity.  In linear response, the lattice displacements are
calculated as a single-valued function of the perturbing force, and
the tweed would therefore form and fade without history dependence.
The temperature dependence of the tweed pattern would arise from the
(single--valued) temperature dependence of the elastic constants, and
it is difficult to conceive of a possible source of hysteresis in this
mechanism.

Having given the above justification for the $\eta \phi^2$ coupling
which we have investigated in detail, we would like to go on to
consider other possible couplings.

\noindent{\bf Coupling to Order Parameter Linearly:}
The system in which tweed has received the greatest amount of
attention in the last several years is the high temperature
superconductors.  In the YBCO type materials, there are twice as many
oxygen sites in the Cu-O planes as oxygen atoms, and the tetragonal to
orthorhombic martensitic transition occurs as the randomly distributed
oxygen atoms break the twofold symmetry between sites and
preferentially align along one axis.  This alignment results in an
increased lattice constant in one direction relative to the other, and
a net rectangular deformation in the Cu-O plane.  In this case, if the
oxygen distribution is taken as the disorder field, disorder clearly
couples directly to $\phi$, and the corresponding term in a Landau
free energy describing this system would appear as a term linear in
disorder and strain, $\eta \phi$.  Even though the coupling is linear,
it is conceivable that hysteresis may arise from the complicated
dynamics of the diffusing oxygen.

This model has been extensively studied.  Semenovskaya {\it et
al.}\refto{Khachaturyan, Khachaturyan2} and Parlinski {\it et
al.}\refto{Parlinski} have each considered a model for YBCO in which
diffusing oxygen atoms arrange into martensitic microdomains, and the
coupling between oxygen position and elastic strain leads directly to
an unquestionably tweedy lattice deformation.  Morphologically, the
tweed structures found in these studies is essentially identical to
that presented in this paper; not only is the qualitative appearance
identical, but quantitatively the correlation lengths are comparable
as well.  However, there is a fundamental difference in the {\it
physical nature} of the tweed structures.  In their investigations,
the tweed is a not an equilibrium phase, but rather a non-equilibrium
or metastable configuration.  Semenovskaya {\it et
al.}\refto{Khachaturyan} observe tweed as a intermediate structure as
the oxygen distribution gradually evolves through an ordering process,
passing through a transient stage (or getting stuck in a metastable
well) consisting of highly anisotropic microdomains before ultimately
reaching the equilibrium twinned martensitic configuration.  In
addition to this transient tweed, Parlinski {\it et al.} also find
``embryonic'' tweed resulting from thermal fluctuations {\it above}
the transition temperature, which they treat analytically as critical
fluctuations in a second order transition.  These studies have
produced a dynamic or metastable tweed in contrast to the static
equilibrium tweed phase which our model seeks to explain.  Yet,
despite the compelling analogy, it is not quite accurate to infer they
have produced the liquid out of which our glass forms!

The nature of the disorder in their nonequilibrium tweed is
fundamentally different from the nature of the disorder in our
equilibrium tweed.  The distribution of oxygen atoms constitutes their
disorder: randomly scattered or clustered into microdomains or ordered
into martensitic variants, the oxygen atoms couple to the strain and
give rise to some lattice deformation.  Yet, since the oxygen atoms
diffuse in response to external parameters, {\it oxygen is not a
source of quenched-in disorder.} The oxygen is effectively an
additional degree of freedom which the system integrates out as it
searches for a stable equilibrium.  In this light, it is clear why
there is no stable, static, equilibrium tweed phase to be found in
these models.  In contrast, the model presented in this paper relies
on the presence of intrinsic quenched-in disorder, in the form of
static compositional inhomogeneities, in order to stabilize the tweed
phase.

Why then, is static tweed seen in YBCO at all?  Significantly, it is
when YBCO is doped with an impurity that static tweed appears.  After
substituting copper with as little as 1.5\% of a transition metal
element (such as Fe, Co, Al, or Ga), TEM observations of tweed are
made\refto{Zhu}.  These impurity atoms are frozen in at temperatures
well above the tweed regime\refto{Zhu}, typically at $> 700K$, and
therefore the impurity disorder is truly quenched-in.  Studies by
Jiang {\it et al.}\refto{Jiang} and Krekels {\it et
al.}\refto{Krekels} have demonstrated static microdomain formation due
to quenched-in impurity atoms.  In these studies, the oxygen--copper
and oxygen--impurity interactions are such that the impurity atoms,
which prefer nearest-neighbor oxygen occupancy, confound the oxygen
chain alignment preferred by the surrounding matrix of copper
atoms\refto{DeFontaine}.  It is through the compromise arrangement of
the oxygen atoms that the static impurity disorder is communicated to
the elastic deformation.  Recent simulation studies by Semenovskaya
{\it et al.}\refto{Khachaturyan2} have included quenched-in impurities
as well as the long-range strain interactions, and have indeed
generated a static, apparently stable, tweed phase.  Ultimately, in
such a model, the impurity atoms act as sites favoring the square
phase amid copper sites favoring the rectangular phase.  The
compositional variation therefore selects a phase (square or
rectangular) but it does not select a particular martensitic variant:
the symmetry of a copper site within the unit cell prevents it from
coupling linearly to the rectangular strain.  It is just this physical
situation which is represented in our general model by a quadratic
coupling between compositional disorder $\eta$ and rectangular strain
$\phi$.  The present study, which seeks to clarify the general
principle involved in equilibrium tweed, and the studies cited above,
which seek to elaborate in valuable detail the specific mechanism at
work in the case of YBCO, therfore complement each other very well.

\noindent{\bf Coupling to Gradients of Disorder:}
Although the disorder on an atomic scale can be described by a scalar
field, it is possible for the random compositional variations to
conspire to produce clusters of more complex symmetry, which may then
collectively couple to the order parameter.  For example, as Robertson
and Wayman\refto{Tweed_NiAl} have argued with respect to NiAl, random
clusters of nickel atoms have a high probability of having tetragonal
symmetry, and therefore coupling to the martensitic strain.  In the
language of the Landau--Ginzburg formalism, this corresponds to a
coupling between higher order gradients of the disorder field and the
order parameter.

\com{
The term $\phi (\partial_x^2 - \partial_y^2) \eta$ has a very
intuitive physical interpretation, as follows.  Generally, the
effective atomic radii of the two (or more) species in an alloy are
measurably different.  As a result, the lattice will suffer
distortions in response to any disorder in the distribution of the
atoms, contracting around the smaller atoms and expanding around the
larger atoms.  It is easily seen that a strategic arrangement of atoms
can produce a rectangular distortion.  In particular, if the
compositional randomness conspires to provide a given region with an
excess of large atoms relative to its neighboring regions to the
sides, but a deficiency of large atoms relative to the regions above
and below, then that region will be squeezed from above and below, and
expand out toward the sides: i.e.  it will be rectangularly distorted.
The term $(\partial_x^2 - \partial_y^2) \eta$ is a measure of this
precisely this distribution of atoms, and its effect on the order
parameter is represented by the term $\phi (\partial_x^2 -
\partial_y^2) \eta$. }

In two dimensions, the term $\phi (\partial_x^2 - \partial_y^2) \eta$
is a symmetry--allowed coupling between $\eta$ and $\phi$ which is
{\it a priori} no less important than the quadratic coupling we have
used.  Since this term is only linear in $\phi$, it could well be
comparable in magnitude to the quadratic coupling, despite the second
derivative, and be just as effective in generating some lattice
deformation.  The relevant question, however, is: what is the capacity
of such a term to generate {\it tweed}.  This is most easily answered
by reexpressing the term in k-space: $\tilde\phi({\bf k}) (k_x^2 -
k_y^2) \tilde\eta({\bf k})$.  Recall that tweed is correlated along
diagonal directions, and therefore is composed of fourier components
for which $k_x^2 - k_y^2 \rightarrow 0$.  Therefore, although
$(\partial_x^2 - \partial_y^2) \eta$ couples to the order parameter,
it does not do so in a manner which allows it to generate tweed.  (In
three dimensions, a slightly more involved
calculation\refto{KarthaThesis} leads to the same result.)

\noindent{\bf Coupling to Bulk Dilation:}
Perhaps the most commonly considered deformation due to compositional
disorder is that which arises from atomic size mismatch in alloys.
This mechanism is represented by the coupling $\eta e_1$ between
composition and bulk dilation, and although it does
not directly couple $\eta$ and $\phi$, we have repeatedly emphasized
that the strain fields are not independent and are intrinsically
coupled by the compatibility conditions.

In the previous section, we were able to translate the bulk dilation
and diagonal strain contributions to the free energy into non--local
interactions in $\phi$, by integrating those secondary strains out of
the free energy, subject to the compatibility constraint.  We can
carry out the analogous analysis for a term $D \eta e_1$, as
follows:

The variation in the free energy caused by variations $\delta e_1$,
$\delta e_2$, and $\delta \lambda$ is
$$\delta f = (A_1 e_1 + D \eta) \delta e_1+ A_2 e_2 \delta e_2 +
\lambda \Biggl\{\nabla^2 \delta e_1 -  \sqrt 8\ \partial_{xy}
\  \delta e_2\Biggr\}$$
$$+\ \delta \lambda \Biggl\{\nabla^2 e_1 -  \sqrt 8\ \partial_{xy}
\ e_2 - \Bigl(\partial_{xx} - \partial_{yy} \Bigr)
\phi \Biggr\}\eqno(DFvariation)$$
and the corresponding expressions for $e_1$, $e_2$, and $\lambda$ are
$$e_1 = {-1 \over A_1} (\nabla^2 \lambda - D \eta)\eqno(Deofm a)$$
$$e_2 = {\sqrt 8 \over A_2} \Bigl(\partial_{xy}\ \lambda
\Bigr)\eqno(Deofm b)$$
$$ - {1 \over A_1} (\nabla^2)(\nabla^2 \lambda + D \eta) - {8 \over
A_2} \Bigl(\partial_{xxyy} \ \lambda \Bigr) = \Bigl(\partial_{xx} -
\partial_{yy} \Bigl)\phi.\eqno(Deofm c)$$ In k-space this gives us
$$\tilde{\lambda}({\bf k}) = {(k_x^2 - k_y^2)\tilde\phi({\bf k}) +
(k_x^2 + k_y^2) D \tilde\eta({\bf k}) /A_1
\over (k_x^2 + k_y^2)^2/A_1 + 8 (k_x^2 k_y^2)/A_1}
\eqno(Dkspace a)$$

$$\tilde{e_1}({\bf k}) = {(k_x^2 + k_y^2) (k_x^2 -
k_y^2)\tilde\phi({\bf k})/A_1 - (k_x^2k_y^2) D \tilde\eta({\bf k})
/A_1 A_2
\over  (k_x^2 + k_y^2)^2/A_1  + 8 (k_x^2 k_y^2)/A_2}\eqno(Dkspace b)$$

$$\tilde{e_2}({\bf k}) = {-\sqrt 8 (k_x k_y) (k_x^2 -
k_y^2)\tilde\phi({\bf k})/A_2 - (k_x k_y)(k_x^2 + k_y^2)D
\tilde\eta({\bf k}) /A_1 A_2
\over (k_x^2 + k_y^2)^2/A_1 + 8 (k_x^2 k_y^2)/A_2}
\eqno(Dkspace c)$$

Again, the free energy can now be written in terms of $\phi$ alone.
The contributions due to $A_1\, e_1^2$ and $A_2\, e_1^2$ are identical
to those found in in the previous section, as they must be, with the
terms proportionate to $\tilde\eta$ cancelling.  The contribution from
the disorder term, $D \eta e_1$ is now a non-local interaction between
disorder and $\phi$ (where the term quadratic in disorder can be
dropped by redefining the zero of the energy): $$F_\eta = \int\!\!\int
d{\bf r}\ d{\bf r'}\ \ f_\eta({\bf r}, {\bf r'})\ \phi({\bf r})\
\eta({\bf r'})
\eqno(nonlocaldisorder)$$ where $$f_\eta = \int {d{\bf k} \over a} {D
\over A_1} {(k_x^2 + k_y^2)(k_x^2 - k_y^2) \over (k_x^2 + k_y^2)^2/A_1
+ 8 (k_x^2 k_y^2)/A_2} e^{i {\bf k} \cdot ({\bf r} - {\bf
r'})}.\eqno(nonlocalintdisorder)$$ More transparently, we can write
$$F_\eta = \int {d{\bf k} \over a} {D \over A_1} {(k_x^2 +
k_y^2)(k_x^2 - k_y^2) \over (k_x^2 + k_y^2)^2/A_1 + 8 (k_x^2
k_y^2)/A_2}
\tilde\phi({\bf k})\ \tilde\eta(-{\bf k}).\eqno(nonlocalintdisorderK)$$

In the previous section we casually argued that a term like this one
will be ineffective in generating a tweedy deformation because it
vanishes precisely for those Fourier modes $(k_x \simeq \pm k_y)$
which correspond to tweed.  A more careful analysis would account for
the fact that the restoring force $A_\phi({\bf k})$ is ${\bf
k}$--dependent, and diminishes as $k_x \pm k_y \rightarrow 0$, {\it
enhancing} the effectiveness of this driving force.  The term in the
last section turns out to indeed be harmless, but here we give a more
complete analysis.  Within linear elasticity, it is possible to solve
explicitly for the deformation resulting from Eq.
\(nonlocalintdisorderK).

We are considering the harmonic free energy $$F = \int {d{\bf k} \over
a} {A_\phi({\bf k}) \over 2} \, |\phi({\bf k})|^2 + D\, Q_1({\bf k})\,
\tilde\phi({\bf k})\,
\tilde\eta(-{\bf k})\eqno(linearfree)$$
where $A_\phi({\bf k})$ and $Q_1({\bf k})$ defined in Eqs.
\(renormalizedc) and \(kdep a) respectively.  This is minimized when
$$\tilde\phi({\bf k}) = {-D\, Q_1({\bf k})\, \tilde \eta(-{\bf k}) \over
A_\phi({\bf k})}.\eqno(phimin)$$ The most revealing measure of this
simple result is the corresponding diffraction pattern for a white
disorder distribution $\eta({\bf k}) =
\eta$, which we calculate using
$$\pmatrix {\tilde\phi({\bf k}) \cr \sqrt 2\, \tilde e_2({\bf k})} =
{1 \over \sqrt 2} \pmatrix{k_x & -k_y \cr k_y & k_x}
\pmatrix{U_x({\bf k}) \cr  U_y({\bf k})}\eqno(strainsolve b)$$
and $$\tilde e_2({\bf k}) = -Q_2({\bf k}) \tilde \phi({\bf k}) - {D
\eta\over A_1} Q_3({\bf k})\eqno(e2is)$$ where $$ Q_3({\bf k}) = {(k_x
k_y)(k_x^2 + k_y^2)/A_2
\over (k_x^2 + k_y^2)^2/A_1 + 8 (k_x^2 k_y^2)/A_2}.\eqno(kdepz)$$
We can invert Eqs. \(strainsolve) and use Eqs. \(phimin) and \(e2is)
to find $$\pmatrix{ U_x({\bf k}) \cr U_y({\bf k})} = {\sqrt 2 \over
k^2} \pmatrix{k_x & k_y \cr -k_y & k_x}
\pmatrix{ -D\eta \, Q_1({\bf k}) \over A_\phi({\bf k}) \cr \cr \cr
\sqrt 2 \biggl\{ {D\eta \, Q_1({\bf k})\, Q_2({\bf k}) \over
A_\phi({\bf k})} - {D \eta \over A_1} Q_3({\bf k})\biggr\} }\eqno(solveU)$$


The scattering contours in figure~11 were calculated
from the solution \(solveU) to the linear problem.  For finite
anisotropy, the diffuse scattering deviates substantially from that
expected from tweed, showing that a substantial amount of non--tweedy
deformation is occurring in the lattice.

This is in agreement with the much earlier analysis of Cochran and
Kartha\refto{Dad} who show that the long--range strain field
associated with random variation in bulk dilation will lead to diffuse
scattering with a strong radial component at $\langle 0 h \rangle$
Bragg peaks.  This is qualitatively distinct from the diagonal diffuse
streaking seen in tweed.  As the anisotropy increases however, the
diffuse scattering lobes converge toward the diagonal directions, and
grow increasingly similar to the diagonal streaks associated with
tweed.

This suggests that within linear elasticity a coupling $\eta e_1$
between disorder and bulk dilation is ineffective at generating tweed
for most realistic materials parameters.  As a more fair test of this
coupling, we must also determine whether tweed might still appear in
the full non-linear model that includes terms anharmonic in the $\phi$
strain. Of course, an analytical solution is now unfortunately
inaccessible, but introducing the coupling to bulk dilation into our
numerical simulation with finite anisotropy, we've determined that
this coupling does not result in a identifiably tweedy modulation.
The configuration in Figure~12 was obtained by eliminating the
coupling $\eta \phi^2$ and replacing it with a coupling $\eta e_1$ of
the same strength.  The lattice is noticeably deformed, yet the
modulations clearly do not constitute tweed.


\bigskip
\noindent{\bf Conclusion}\hfil\break
We have reported the results of a simulation of the pretransitional
behavior of a two dimensional martensitic model material.  Unlike the
historically traditional procedure of seeking and studying in local
detail the coupling of special localized defects to first order
transitions, we show that intrinsic statistical variation of
composition suffices to generate tweed patterns, as well as to provide
an understanding of how a tweed region can exist in a transitional
regime over an extended temperature range above the nominal bulk
transformation temperature.  Most importantly, the model described
here has uncovered a tweed that is far more complex than a mere
lattice response to local defects.  Long--range, cooperative,
non-linear processes give rise to tweed which is, in the infinite
anisotropy limit, a distinct stable thermodynamic phase between the
austenite and martensite phases, and which is moreover a glass phase
that exhibits properties distinct from the phases it separates: slow
relaxation, a diverging non-linear susceptibility, glassy dynamics.
In actual materials which do not have infinite anisotropy, the phase
transitions bounding the glassy phase will be rounded, but many
real--world materials are indeed anisotropic enough that the tweed
regime is likely to still exhibit experimentally observable glassy
behavior.

{}From the modeling point of view we have once again demonstrated the
utility of nonlinear, nonlocal free energy functions, now extended to
non-uniform, disordered systems, for representing the mesoscale
phenomenology of patterns in lattice distortive phase transitions.
This formalism makes contact with a wide class of other displacive
transitions in metals, non-metals and ceramics, e.g. ferroelectricity,
ferroelasticity, perhaps even biomolecular conformation
changes\refto{BruceCowley}.  It has been our aim in this paper to
outline the method in sufficient detail that it can now be applied to
various materials, at least semi-quantitatively.

\com{
The current form of this model has some substantial limitations which
must be acknowledged.  First, temperature is incorporated in only a
limited way, by using empirical materials parameters which are
temperature dependent.  This allows us to take into account effects
such as the softening of $A_\phi$, or, say, thermal expansion, (by
introducing a temperature dependent lattice constant).  However, it is
still a zero temperature model: there is no {\it dynamics} in the
model and thermal fluctuations are neglected.  Since tweed is observed
even above room temperature, thermal activity is undoubtedly
important: the thermal vibration in a real lattice is substantial, and
displacements from equilibrium positions become comparable in size to
the displacements involved in not only the tweed modulation, but the
martensitic deformation as well.  Temperature related effects must
have some bearing on the stability and structure of the tweed that is
observed in real materials.  Work in progress suggests that
introducing thermal motion into the model will quantitatively, but not
qualitatively, affect the phase diagram.  There are two relevant
competing effects: 1) finite temperatures will wash out local
distortions having energies which are very small compared to $k_B T$,
but 2) the greater entropy of the highly disordered tweed
configuration relative to the rather uniform austenite will be favored
by finite temperatures.  (The second point has been addressed by
Fujita, who proposed that tweed could be stabilized by the entropy of
mixing of the regions of different phases\refto{Fujita}.)
}

A logical continuation of the present research would, of course,
extend this model and simulation to three dimensions.  Like many
attempts to address pretransitional mesoscale modulations, this
approach casually seeks to investigate a three dimensional phenomena
with a two dimensional model, and various problems arise with this
uncontrolled approximation.  A two dimensional system will certainly
be more unstable toward a lattice distortion such as tweed than a real
three dimensional material.  (Indeed, in two dimensions there is
generally {\it no} stable finite temperature crystalline phase.)  Also, in
attempting to describe a three dimensional material with a two
dimensional model, one must use materials parameters from the real
material, and hope that the model will yield similar behavior, but it
is possible that qualitatively different physics may be important in
the real 3D sample.  For example, in modeling planar compounds such as
the high $T_c$ materials, such a two dimensional approach still
incorporates the appropriate physical symmetries.  On the other hand,
cubic materials have additional compatibility conditions and the limit
of infinite elastic anisotropy yields a six component Potts-like model
rather than an Ising model\refto{KarthaThesis}.  We have yet to
investigate how this may cause tweed in cubic materials to differ from
tweed in tetragonal materials.

An additional issue deserving further research is the softening
behavior in the pretransitional tweed regime.  We have long suspected
that the pretransitional ``anomalies'' in elastic constant behavior
are inextricably connected to the presence of tweed type modulations.
Whereas conventional wisdom holds that elastic softening leads to
pretransitional modulations (and ultimately to the martensitic
transformation), we believe that the softening can in turn be {\it
enhanced} by the pretransitional modulation.  A bulk elastic constant,
measured over an entire macroscopic specimen, will necessarily reflect
not simply the harmonic response arising from the bare interatomic
potentials, but also the mesoscopic lattice response to the applied
driving force.  When stressed, a modulation such as tweed will
certainly respond elastically, but it will also flip domains, depin
boundaries, rearrange clusters etc.  We believe that it is precisely
such nonlinear processes which account for much of the behavior
underlying the pretransitional anomalies in elastic softening,
internal friction, and acoustic attenuation.  We've undertaken to
investigate these effects in our model by studying the response of a
simulated patch of tweed to an externally applied strain.  Allowing
these relaxational processes, we measure elastic constants which are
substantially softer that the ``bare'' elastic constants otherwise
found.  Furthermore, we observe the dissipative and hysteretic effects
which underlie the experimental ultrasonic attenuation and internal
friction measurements\refto{KarthaThesis}.

Many of the ideas which have been incorporated into this model have
also been of central importance in earlier work by a number of other
investigators.  Ericksen\refto{Ericksen} and Jacobs\refto{Jacobs} have
considered the limit of infinite anisotropy and shown the general form
for allowed solutions.  Semenovskaya {\it et al.} and Parlinski {\it et
al.} have noted the vital importance of the long-range nature of
strain fields in a lattice modulation such as
transient\refto{Khachaturyan} and dynamical\refto{Parlinski} tweed.
Jiang {\it et al.}\refto{Moss}, Krekels {\it et al.}\refto{Krekels} and
Becquart {\it et al.}\refto{Becquart} have recognized the importance of
compositional randomness (i.e. random placement of alloy components or dopants)
in
determining the tweed structure.  We have assembled these various
ingredients into a strikingly simple and powerfully general model
which provides compelling answers to two questions which have been
troubling investigators of pretransitional phenomena in martensitic
materials for many years: What is tweed? Why does it occur?  Moreover,
the answers given here have established an unexpected connection
between tweed and spin glasses, suggesting a new line of experimental
investigation into tweed.  The signature of glassiness may be
observable in tweed through measurements of nonlinear elastic
constants at the onset of tweed, measurements of frequency dependent
relaxation phenomena using ultrasonic attenuation, or investigations
of ``remanent strain'' through hysteresis measurements, for example.
In establishing this connection between tweed and glasses, it is hoped
that the tools being developed within the field of disordered systems
in condensed matter theory may be brought to bear on the problem of
pretransitional phenomena in martensitic systems.

We would like to thank Lee Tanner for helpful and enjoyable
discussions, and for providing the TEM image used in Figure~1.  We
also thank Teresa Cast\'an, who was a very welcome collaborator in the
spin glass aspects of this research.  We acknowledge the support of
DoE Grant \#DE-FG02-88-ER45364.  SK also acknowledges support from the
Department of Education and from the Sloan Foundation, Grant \#93-6-6.

\com{
In addition, however, we want to note that the role of statistical
composition variations in inducing qualitatively new phenomena is not
an isolated event.  Increasingly, key properties of real--world
materials are indeed being attributed to the presence of disorder.
Levitation in type II superconductors relies on flux lines sticking at
inhomogeneities; current--voltage characteristics of charge density
waves is shaped by pinning at impurities; dynamics of coarsening and
domain growth is altered in the presence of quenched randomness;
localization effects in semiconductors appear in the presence of
doping; and hysteretic behavior of magnetic systems is determined by
the nature of the intrinsic disorder\refto{NATOGeilo}-- to list but a
few examples.}

\noindent
\nobreak

\references

\refis{Falk} F. Falk, {\sl Z. Phys. B - Condensed Matter}, {\bf 51},
(1983), 177.

\refis{Jacobs} A. E. Jacobs, {\sl Phys. Rev. B}, {\bf 31}, (1985), 5984.

\refis{Tweed_NiAl}
I. M. Robertson and C. M. Wayman, {\sl Phil. Mag. A}, {\bf 48}, (1983)
421, 443, 629.  D. Schryvers and L. E. Tanner, {\sl Ultramicroscopy},
{\bf 37}, (1990) 241).  L. E. Tanner, D. Schryvers and S. M. Shapiro,
{\sl Mater. Sci. Eng. A}, (1990) 205.

\refis{TannersPicture} Figure 1 appears courtesy of Lee Tanner.

\refis{Tweed_FePd} S. Muto, S. Takeda, R. Oshima, and F. Fujita, {\sl
J. Phys: Cond.  Mat.}, {\bf 1}, (1989) 9971.  S. Muto, S. Takeda and
R. Oshima, {\sl Jap. J. Appl.  Phys.}, {\bf 29}, (1990), 2066.

\refis{Oshima_Hysteresis} R. Oshima, M. Sugiyama, and F. E. Fujita,
{\sl Met. Trans. A}, {\bf 19}, (1988) 803.

\refis{Muto_Softening} S. Muto, R. Oshima, and F. Fujita, {\sl Acta Metall.
Mater.}, {\bf 4}, (1990), 685.

\refis{Sato} M. Sato, B. H. Grier, S. M. Shapiro, and H. Miyajima,
{\sl J. Phys: Met. Phys.}, {\bf 12}, (1982), 2117-2129.

\com{
\refis{Fujita} F. Fujita, {\sl Mat. Sci. and Eng.}, {\bf A127},
(1990), 243.
}

\refis{Tweed_CuAu} K. Yosuda and Y. Kanawa, {\sl Trans. Jap. Inst.
Metals}, {\bf 18}, (1972) 46.

\refis{Tweed_YBCO} W. W. Schmahl, A. Putnis, E. Salje, P. Freeman, A.
Graeme--Barber, R. Jones, K. K. Singer, J. Blunt, P. P. Edwards, J.
Loram, and K. Mirza, {\sl Phil. Mag. Let.}, {\bf 60}, (1989) 241. T.
Krekels, G. van Tendeloo, D. Broddin, S. Amelinckx, L. Tanner, M.
Mehbod, E. Vanlathem, and R.  Deltour, {\sl Physica C}, {\bf 173},
(1991) 361;

\refis{Zhu} Y. Zhu, M. Suenaga and J. Tafto, {\sl Phil. Mag. Let.},
{\bf 62}, (1990), 51; Y. Zhu, M. Suenaga and A. R. Moodenbaugh, {\sl
Phil. Mag. Let.}, {\bf 64}, (1991), 29.

\refis{Moss} X. Jiang, P. Wochner, S. C. Moss, and P. Zschack,
{\sl Phys. Rev. Let.}, {\bf 67}, 2167, (1991).  {\sl Proceedings of
International Conference on Martensitic Transformations}, Monterey,
CA. 20-24 July, 1992.

\refis{Tweed_A15s} T. Onozuko, N. Ohnishi and M. Hirabayahsi, {\sl
Metall. Trans. A+}, {\bf 19}, (1988) 797.

\com{
InTl:
CuZnAl:
LaCuO:
}

\refis{Heuer} A.~H.~Heuer, R.~Chaim,and V.~Lanteri, {\sl Adv. in
Ceramics},{\bf 24}, (1988) 3; A.~H.~Heuer and M.~Rh\"ule, {\sl Acta
Met.}, {\bf 12}, (1985) 2101.

\refis{Tweed_CuBe} L. Tanner, {\sl Phil. Mag.}, {\bf 14}, (1966), 111.

\refis{Tweed_steel} G. R. Speich and K. A. Taylor, in {\sl
Martensite}, eds. G. B. Olson and W. S. Owen, (ASM International,
1992), 243.

\com{
\refis{NATOGeilo} Proceedings of the 22nd NATO Advanced Study
Institute, Geilo, Norway April 13-20, 1993.  }

\refis{Landau} L.D.Landau and D.M.Lifshitz, {\sl Statistical Physics},
(Pergamon, Oxford, 1968), 2nd ed.

\refis{Sokolnikoff} I.S.Sokolnokoff, {\sl Mathematical Theory of
Elasticity}, (McGraw--Hill, New York, 1946).

\refis{Baus} M.Baus and R.Lovett, {\sl Phys. Rev. B}, {\bf 65},
(1990), 1781.

\refis{Imry} The analogous mechanism in the case of the random field
Ising model has been investigated by Y. Imry and M. Wortis, {\sl Phys.
Rev. B}, {\bf 19}, (1979), 3580.

\refis{Petry} A. Heiming, W. Petry, J. Trampenau, M. Alba, C. Herzig,
H. R. Schober, G. Vogl, {\sl Phys. Rev. B}, {\bf 43}, (1991) 10948.
See also parts I and II of this series, {\sl Phys. Rev. B}, {\bf 43},
(1991) 10933 and 10963.

\refis{TannerWuttig} L. E. Tanner, M. Wuttig, {\sl Mat. Sci. and Eng.},
{\bf A127}, (1990) 137.

\refis{BarschKrumhanslinOlson} G. R. Barsch and J. A. Krumhansl, in
{\sl Martensite}, eds. G. B. Olson and W. S. Owen, (ASM International,
1992), 126.

\refis{Gobin} P. F. Gobin, and G. Guenin, in {Solid State Phase
Transformations in Metals and Alloys}, proceedings of Ecole d'\'et\'e
d'Aussois, Sept. 3-15, 1978, (Les \'Editions de Physiques, Orsay,
France)

\refis{Krivoglaz} M. A. Krivoglaz, {\sl Theory of X-Ray and Thermal
Neutron Scattering by Real Crystals}, (Plenum, New York, 1969).

\refis{Kartha} S. Kartha, T. Castan, J. A. Krumhansl, and J. P. Sethna,
{\sl Phys. Rev. Let.}, {\bf 67}, (1991), 3630.  J. P. Sethna, S.
Kartha, T. Cast\'an, and J. A. Krumhansl, {\sl Physica Scripta}, {\bf
T42}, (1992), 214.

\refis{KarthaThesis} S. Kartha, Ph.D. Thesis, Cornell University, 1994,
unpublished.

\refis{KrumhanslGooding} J. A. Krumhansl and R. J. Gooding, {\sl Phys.
Rev. B}, {\bf 39}, (1989), 3047.

\refis{BarschKrumhansl} G. R. Barsch and J. A. Krumhansl, {\sl Met.
Trans. A}, {\bf 19A}, (1988), 761.

\refis{Ericksen} Discussed by J.~L.~Ericksen, {\sl Int'l J. of Solids and
Structures}, {\bf 22}, (1986), 951.

\refis{Parlinski} K.~Parlinski, E.~K.~H.~Salje, and V.~Heine, {\sl
Acta Metall. Mater.}, {\bf 41}, (1993), 839.  K.~Parlinski, V.~Heine,
and E.~K.~H.~Salje, {\sl J. Phys.:Condens. Matter}, {\bf 5}, (1993),
497.  E.~Salje and K.~Parlinski, {\sl Supercond. Sci.  Technol.}, {\bf
4}, (1991), 93.  A.~M. Bratkovsky, S.~C. Marais, V.~Heine, and E.~K.
H.~Salje, {\sl J. Phys.: Condens. Matter}, {\bf 6}, (1994), 3679.

\refis{BruceCowley} A. D. Bruce and R. A. Cowley, {\sl Structural
Phase Transitions}, (Taylor and Francis, London: 1981).

\refis{Becquart} C. S. Becquart, P. C. Clapp, and J. A. Rifkin, {\sl
Phys. Rev. B}, {\bf 48}, (1993), 7.

\refis{EA} This is related to the Edwards-Anderson order parameter in
spin glass literature; see, for example, K. Binder and A. P. Young,
{\it Rev. of Mod. Phys.}, {\bf58}, (1986) 801.

\refis{spinsbeyond} M. Mezard, G. Parisi, and M. A. Virasoro,
 {\it Spin Glass Theory and Beyond}, (World Scientific, Singapore:
1987).

\refis{myexp} The fact that our low temperature correlation data can be
fit to a power law does not contradict this assertion, because careful
spin glass simulations by Ogielski have also observed apparent power
law correlations.  (See {\it Phys. Rev. B}, {\bf32}, (1985), 7384.)
The mean-field theoretical prediction is that correlations will have a
piece which falls off as $t^{-\nu}$ and a static piece which remains
at infinite times. (See H. Sompolinsky and A. Zippelius, {\it Phys.
Rev. B}, {\bf 25}, (1982), 6860.)

\refis{frequency_dependence} M. Wuttig, {\sl Met. Trans. A}, {\bf 19},
(1988) 185.

\refis{LeSar} J. Van Tendeloo, J. Van Landuyt, and S. Amelinckx in
 {\it Competing Interactions and Microstructures: Statics and
Dynamics}, ed. by LeSar, R., A. Bishop, and R. Heffner
(Springer-Verlag, Berlin: 1988)

\refis{Zener} C. Zener, {\sl Phys. Rev},{\bf 71} (1947) 846.

\refis{TannerStatic} L.E. Tanner, S. M. Shapiro, D. Shryvers, and Y. Noda,
{\it Mat. Res. Soc. Symp. Proc.}, {\bf 246}, (1992), 265.

\refis{hardnumber} This partially static tweed had a $\phi$ correlation
of $\xi(t=5000 MCS) = .08$.

\refis{Baus} M.Baus and R.Lovett, {\sl Phys. Rev. B}, {\bf 65},
(1990), 1781.

\refis{Sokolnikoff} I.S.Sokolnokoff, {\sl Mathematical Theory of
Elasticity}, McGraw--Hill, New York (1946).

\refis{Bradley} A. Bradley and A. Taylor, {\sl Proc. R. Soc. A.}, {\bf
159}, (1937), 56.

\refis{Zhu} Y.~Zhu, M.~Suenaga, and J.~Tafto, {\sl Phil. Mag.
Let.}, {\bf 64}, (1991), 29.  Y.~Zhu, M.~Suenaga, and
A.~R.~Moodenbaugh, {\sl Phil. Mag.  Let.}, {\bf 62}, (1990), 51.
Y.~Xu, M.~Suenaga, J.~Tafto, R.~L.~Sabatini, and A.~R.~Moodenbaugh,
{\sl Phys. Rev. B}, {\bf 39}, (1989), 6667.

\com{
\refis{Yamada} H. Seto, Y. Noda, and Y. Yamada, {\sl J. Phys. Soc.
Jap.}, {\bf 59}, (1990), 965.
}

\refis{Khachaturyan} S.~Semenovskaya and A.~G.~Khachaturyan, {\sl
Phys. Rev. Let.}, {\bf 67}, (1991), 2223.  L.-Q.~Chen, Y.~Wang, and
A.~G.~Khachaturyan, {\sl Phil. Mag. Let.}, {\bf 65}, (1992), 15.
S.~Semenovskaya, and A.~G. Khachaturyan, {\sl Physica D}, {\bf 66},
(1993), 205.

\refis{Khachaturyan2} S.~Semenovskaya, and A.~G. Khachaturyan, {\sl
Phys. Rev. B}, {\bf 47}, (1993), 12182.

\refis{Krekels} T.~Krekels, G.~van Tendeloo, D. Broddin, S. Amelinckx,
L. Tanner, M. Mehbod, E. Vanlathem, and R. Deltour, {\sl Physica C},
{\bf 173}, (1991), 361.

\refis{Jiang} X. Jiang, P. Wochner, S. C. Moss, and P. Zschak,
preprint. To appear in {\sl Proceedings of ICOMAT-92}, (Monterey,
California, July 20-24, 1992).

\refis{DeFontaine} D.~De~Fontaine, L.~T.~Wille, and S.~C.~Moss, {\sl
Phys. Rev. B}, {\bf 36}, (1987), 5709.

\refis{Dad} W.~Cochran and G.~Kartha, {\sl Acta Crystallogr.}, {\bf
9}, (1956), 259, 941, 944.

\refis{properferro}
This model is appropriate to so--called ``proper'' ferroelastics, in
contrast to improper ferroelastics where some ``shuffle'' or optical
mode distortion drives the transition\refto{BarschKrumhansl}.

\refis{partialsoftening}
It is important to note that this softening is not a complete
softening to zero frequency.  The sixth order expression for the
energy of the $\phi$ strain gives a triple well potential where the
wells at $\phi = \pm \phi_M$ become the thermodynamically stable phase
when the quadratic coefficient is {\it still positive}, i.e. this is a
first--order transformation and occurs with only a partial softening
of $A_\phi$\refto{BarschKrumhanslinOlson, KrumhanslGooding}.  Indeed
some ferroelastic transitions are true soft mode transitions, but then
these are truly second order and the attendent critical fluctuations
will naturally give rise to some dynamical pretransitional effects.

\refis{straingradcoeff}
A positive strain gradient coefficient will contribute upward
curvature to the dispersion of the related phonon
branch\refto{BarschKrumhansl, KarthaThesis} (in this case the TA$_1$
branch).  Measuring the curvature in experimental data\refto{Sato}
(and accounting for the negative curvature of arising simply from
lattice discreteness) yields the strain gradient coefficient we need.

\refis{phasediagram}
Notice that the horizontal axis is not simply the average composition,
as is usually the case for binary alloy phase diagrams.  Over the
range of interest, composition has a simple relationship to
transformation temperature, and so no additional information would be
revealed.  The strength of the coupling reveals more information, and
is a more appropriate comparison to previous analytical work, as will
be discussed below.

\refis{twinsform}
The twins form for precisely the same reason that domains form in
ferromagnets: elastic strain energy (magnetic field energy) is
minimized by setting up alternating martensitic variants (magnetic
domains) which partially cancel each other's long--range strain field
(magnetic field).

\refis{datafrom}
The top set of results in this figure uses the same data which gave
long time correlations in Figure~4.

\refis{guaranteed}
The computer simulation uses the displacement fields as the degrees of
freedom, so this geometric compatibility is automatically guaranteed;
it is taken into account implicitly because the simulation algorithm
considers distortions of a defect-free lattice.

\endreferences

\vfill\eject

\topinsert\oneandahalfspace
\centerline{\psfig{figure=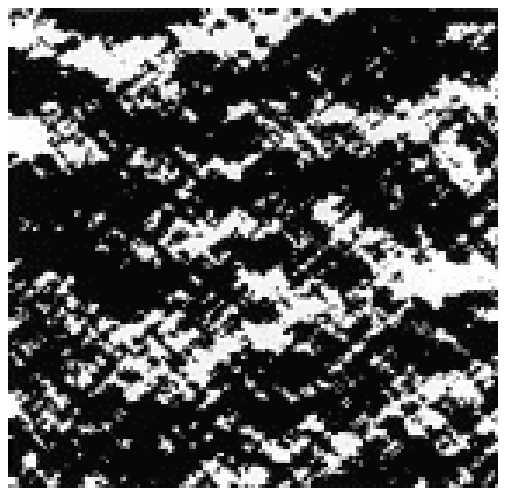,width=4.0truein}}
\medskip \narrower \parskip 0pt Figure~1.  {\bf Tweed}: is shown here
as experimentally observed in transmission
electron microscopy of NiAl.  It is identified by its diagonal
striations, which reflect some pseudo--periodic lattice deformation
with correlations on the scale of some tens of atomic spacings.  This
pattern is consistent with simulation results below.
\endinsert

\topinsert\oneandahalfspace
\vskip.5truein
\smallskip
 \medskip \narrower \parskip 0pt Figure~2(a).  {\bf Simulation Results}:
These configurations are generated by a Monte Carlo computer
simulation based on the continuum elasticity model of a system
undergoing a square $\rightarrow$ rectangular martensitic
transformation, where the transformation strain has been coupled to a
disordered composition field.  The shading reflects the strain order
parameter $\phi({\bf x})$, varying from dark to light as the
strain goes from the horizontally stretched rectangular martensite
variant, to undeformed square phase, to the vertically stretched
variant.  (a) The undeformed austenite phase.  (b) Mottled texture. (c)
Fine tweed. (d) Coarse tweed. (e) Twinned martensite.
\vfill\eject\
\endinsert

\topinsert\oneandahalfspace
\vskip.3truein
\smallskip
\narrower \parskip 0pt Figure~2(b,c).  {\bf Simulation Results.}
\vfill\eject
\endinsert

\topinsert\oneandahalfspace
\vskip.3truein
\smallskip
\narrower \parskip 0pt Figure~2(d,e).  {\bf Simulation Results.}
\vfill\eject
\endinsert

\topinsert\oneandahalfspace
\centerline{\psfig{figure=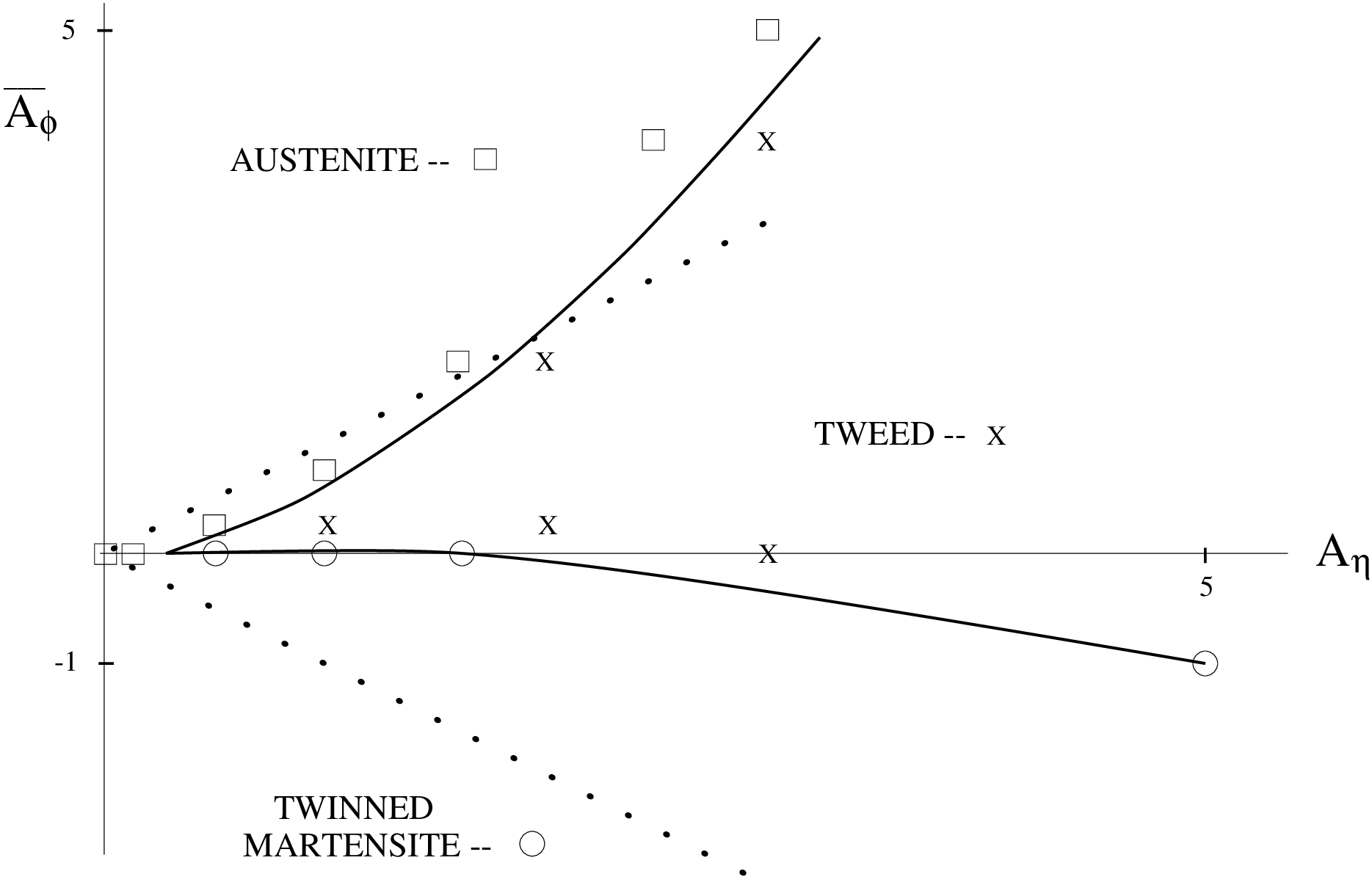,width=6truein}}
\medskip \narrower \parskip 0pt Figure~3.  {\bf Phase Diagram} for
this model is plotted against parameters $\bar{A_\phi}$ and $A_\eta$
(given in units of $10^{10}$ N/m$^2$).  The symbols $\square$, X, and
$\circ$ mark some points in parameter space where the numerically
determined groundstate configuration is austenite, tweed, or
martensite, respectively.  The solid lines are drawn to separate the
resulting three regimes: 1) The AUSTENITE phase is the relatively
undeformed lattice.  2) The TWEED structure develops as a response to
the compositional disorder.  The degree of deformation depends on the
degree of softening and the strength of the coupling to the disorder
field.  3) The TWINNED MARTENSITE is the conventional low temperature
phase.  The dotted lines correspond to the phase boundaries in the
infinite anisotropy approximation, in which the martensitic tweed
problem is mapped to a spin glass\refto{Kartha}.
\endinsert

\topinsert\vskip1.0in
\centerline{\psfig{figure=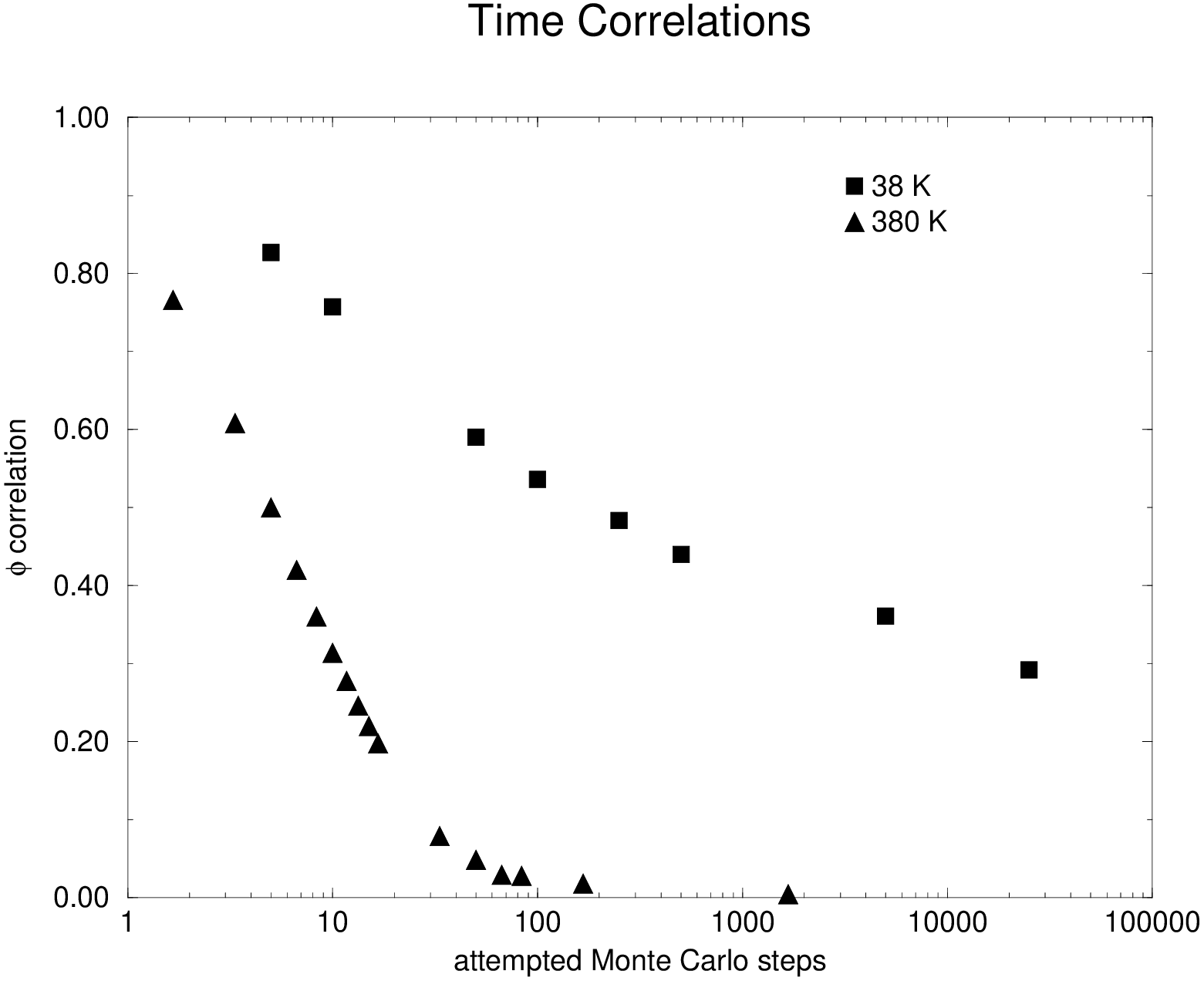,width=4.0in}}
\smallskip
\medskip \narrower \parskip 0pt Figure~4.
{\bf Time Correlations}: Correlations in the $\phi$ martensitic
distortion as a function of time, (where time is measured in attempted
Monte Carlo steps). The physical temperature for each curve is
converted from the corresponding Monte Carlo temperature.  The upper
curve reveals persistent correlations, suggestive of sluggish, glassy
dynamics.
\vfill\eject
\endinsert

\topinsert\oneandahalfspace
\vskip1truein
\bigskip \narrower \parskip 0pt
Figure~5.  {\bf Diffuse Streaking:} around three bragg points is
shown. (a) Experimental x-ray scattering data\refto{Moss} for
YBa$_2$Cu(Al)$_3$O$_{7-\delta}$ around the indicated Bragg peaks.  (b)
Corresponding diffraction data extracted from the computer simulation
of tweed (using FePd parameters) faithfully reproduce important
features of the experimental data, i.e. the diffuse streaking is
highly anisotropic, most pronounced in the $\langle 11
\rangle$ directions, and asymmetrically depends on the Bragg peak
index.
\endinsert

\topinsert
\medskip \noindent \narrower \parskip 0pt Figure~6. {\bf Effects of
Time Averaging for ``Static'' and ``Dynamic'' Tweed}: (a) Instantaneous
snapshot of real-space positions. Notice that the tweed easily
discerned in the upper sample is obscured by the large thermal
fluctuations in the higher temperature (bottom) sample. (b) Average of
30 real-space configurations recorded during simulation. Tweed is
visible in the top row, but the deformation has averaged to almost
zero in bottom row. (c) Average of 30 diffraction patterns calculated
during simulation. Note that instantaneous tweed-like fluctuations are
present in {\it both} cases.  Length of simulations: 150,000 attempted
Monte Carlo steps (approximately one nanosecond).
\endinsert

\topinsert\oneandahalfspace
\centerline{\psfig{figure=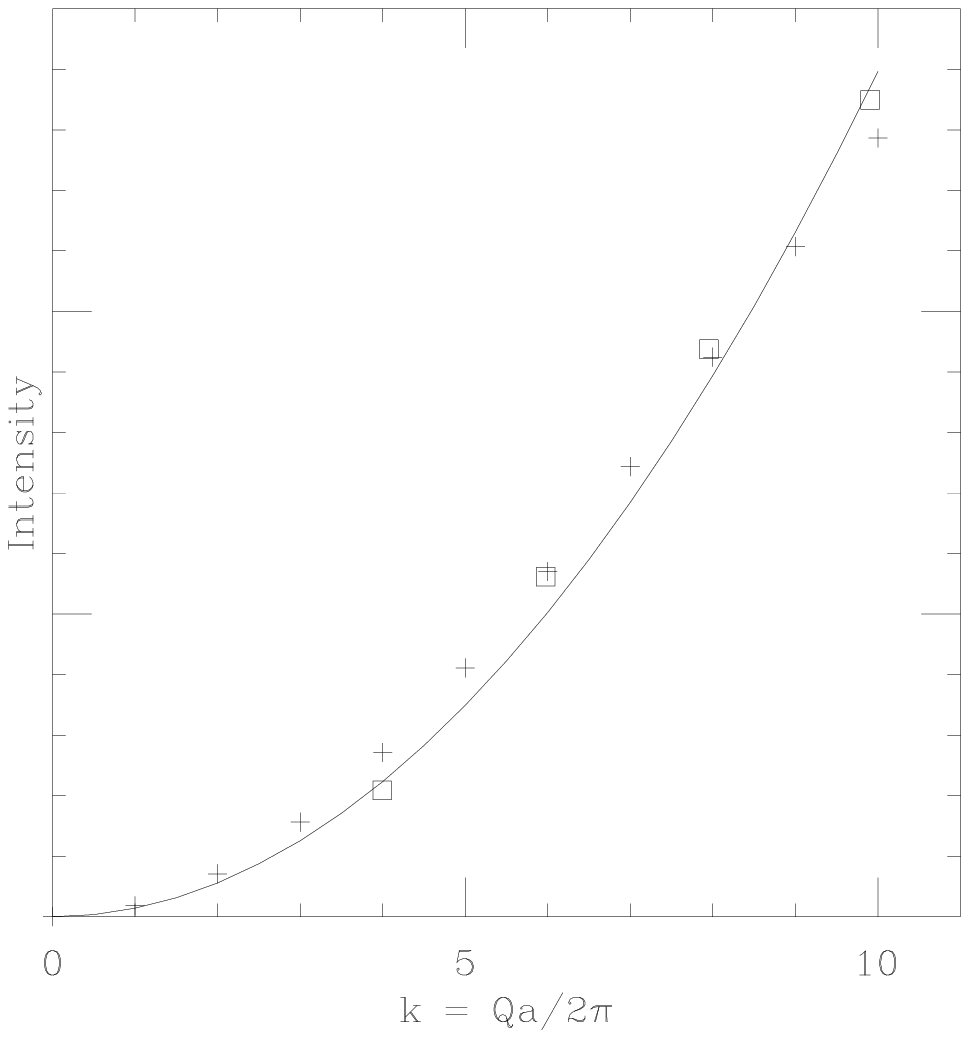,width=4truein} }
\medskip \narrower \parskip 0pt
Figure~7.  {\bf Diffuse Scattering Intensity vs. Wavevector:} The
squares are experimental measurements\refto{Moss} of diffuse
scattering intensity at ${\bf Q} = \langle 0\ Q\ 0 \rangle + {\bf
\epsilon}$ where $Q/(2 \pi / a) = 4,6,8,10$ and ${\bf \epsilon} =
\langle .06\ .06\ 0\rangle$.  The crosses are simulation data,
scaled by a single constant for comparison to the experimental data.
The curve is a fit to $I \sim |{\bf Q}|^2$, which would be exact in
the limit of infinitesimally small displacements.
\endinsert

\topinsert\oneandahalfspace
\centerline{\psfig{figure=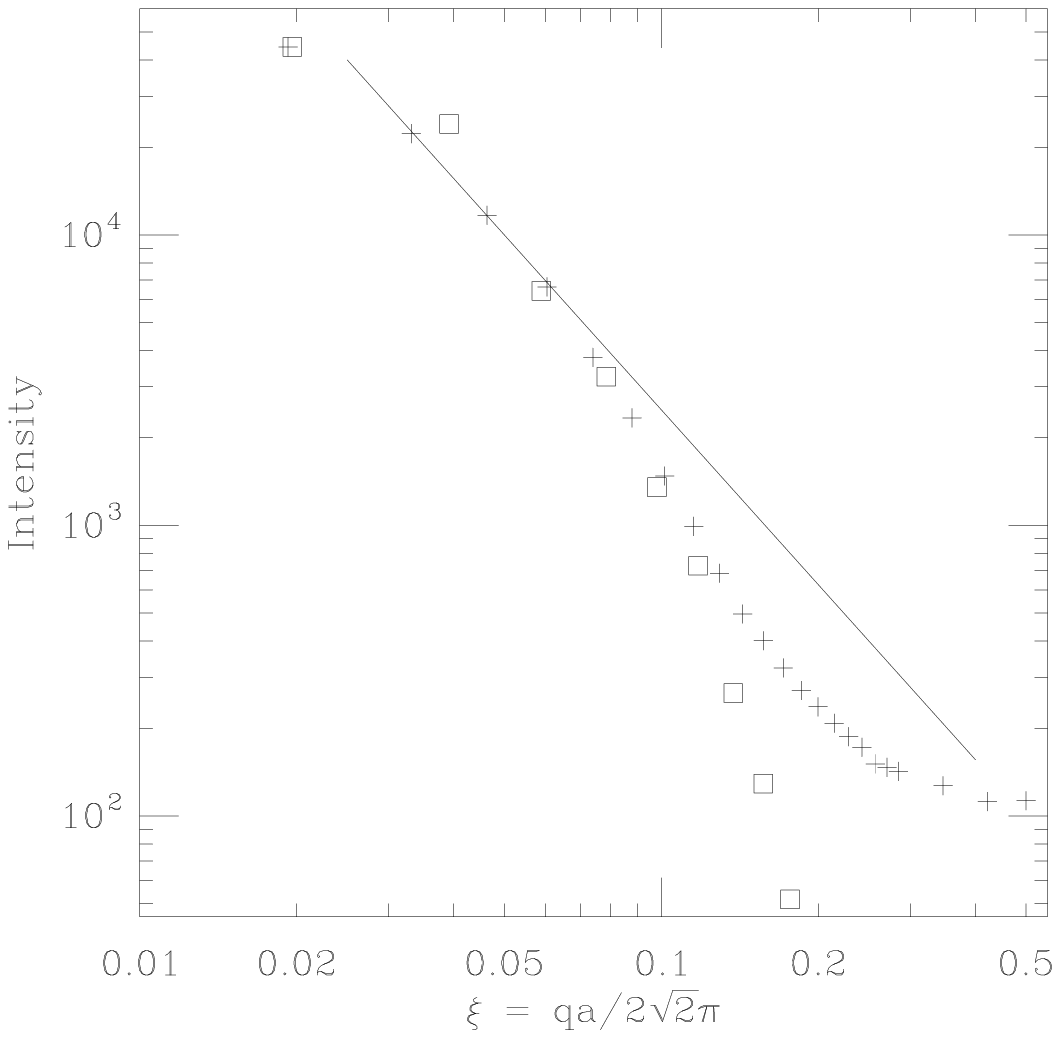,width=4truein} }
\medskip \narrower \parskip 0pt
Figure~8.  {\bf Intensity of Diffuse Scattering} in the $\langle \bar
1 1 0\rangle$ direction around Bragg peak $\langle 660 \rangle$.  The
experimental data\refto{Moss} are shown with crosses, the
simulated data with squares.  The line is for comparison to a
$1/q^2$ dependence.
\endinsert

\topinsert\oneandahalfspace
\centerline{\psfig{figure=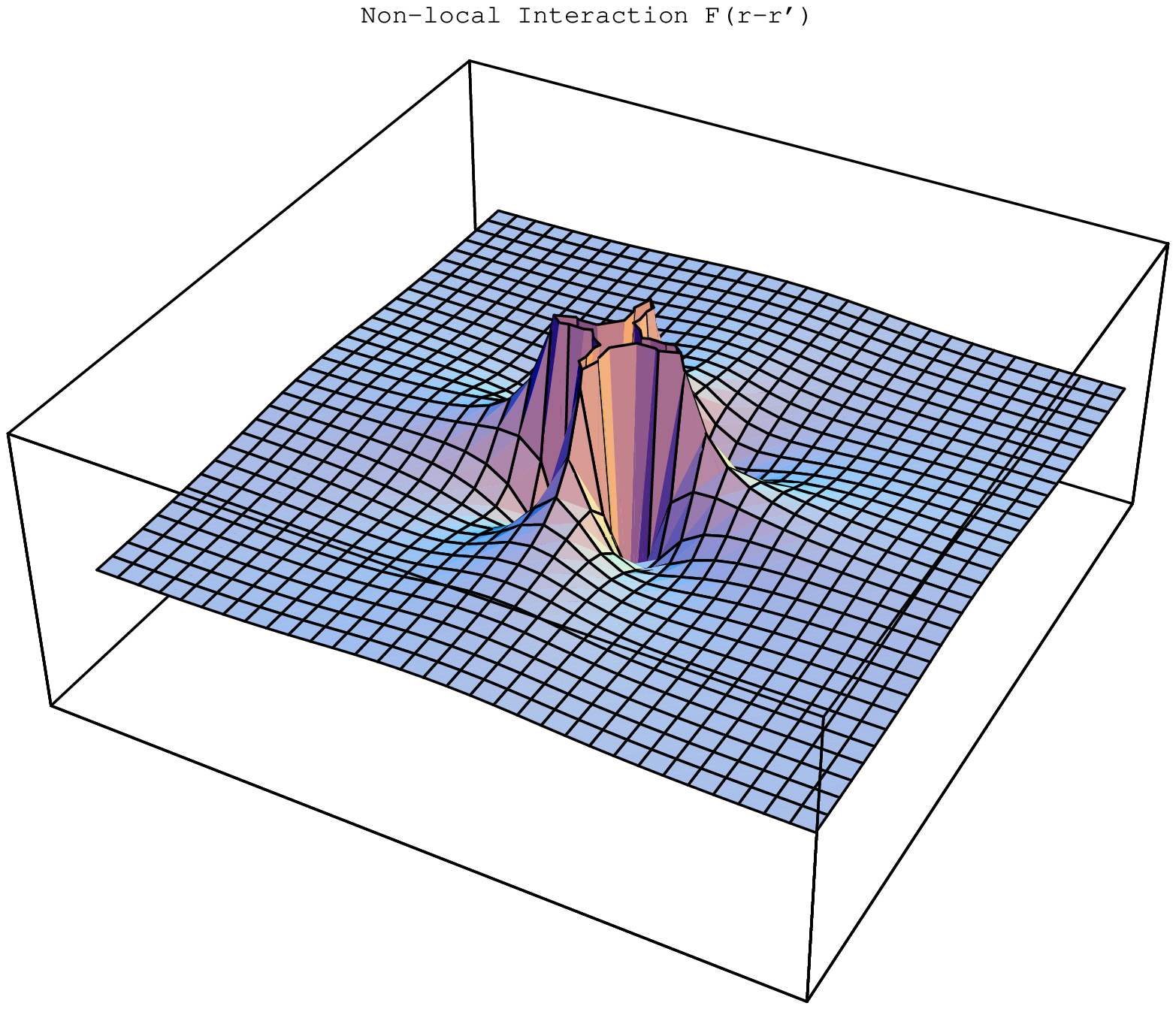,width=4truein} }
\medskip \narrower \parskip 0pt
Figure~9.  {\bf The long--range nonlocal interaction} $f({\bf r},{\bf
r'})$.  Summing the two interactions $f_1({\bf r},{\bf r'})$ and
$f_2({\bf r},{\bf r'})$, and using materials parameters for FePd as in
the simulation, gives us the full form for the nonlocal interaction,
$f({\bf r}-{\bf r'})$ relating $\phi({\bf r})$ and $\phi({\bf r'})$.
(The function is truncated near the origin to maintain a reasonable
scale.)
\endinsert

\topinsert\oneandahalfspace
\centerline{\psfig{figure=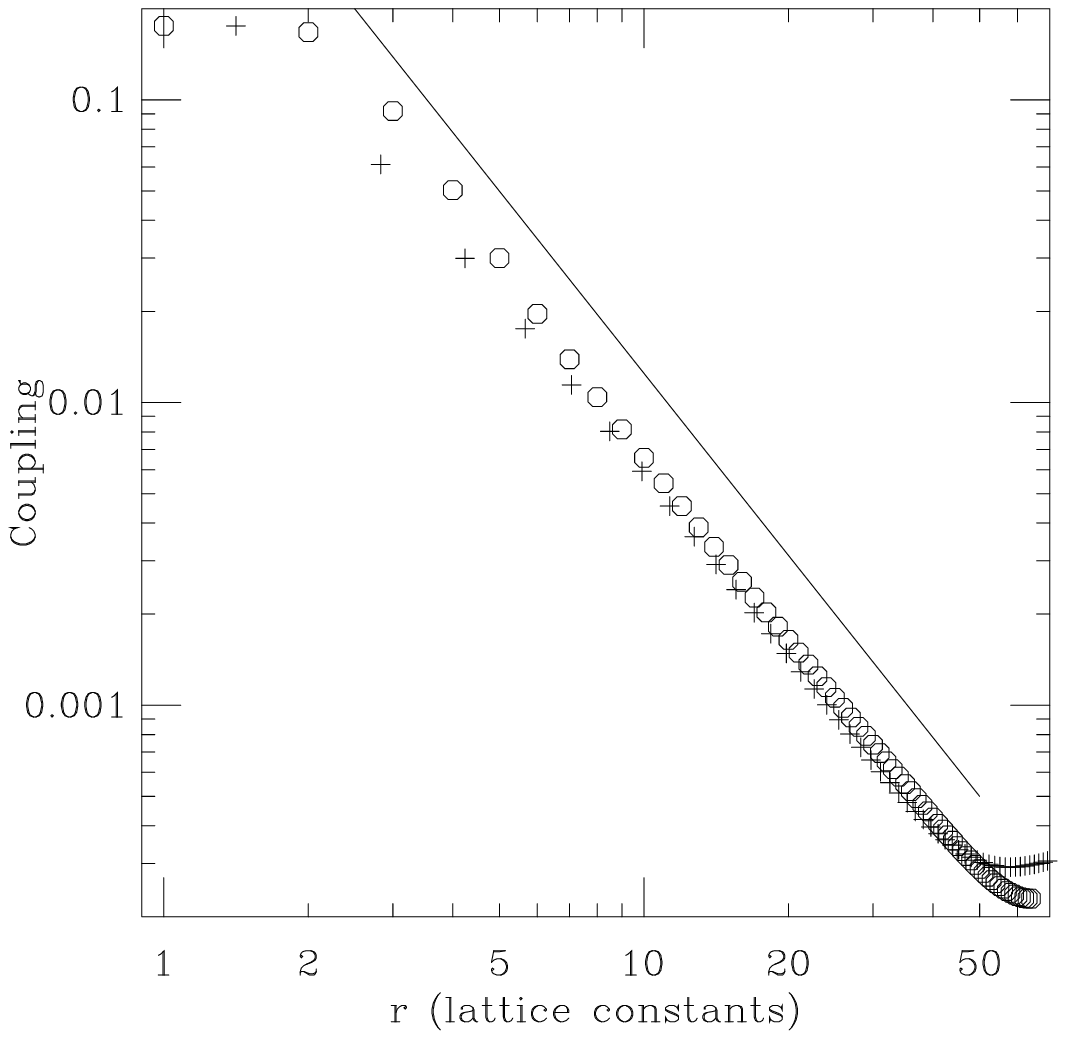,width=4truein} }
\medskip \narrower \parskip 0pt
Figure~10.  {\bf The long--range nonlocal interaction} is plotted
along the diagonal (crosses) and axial (circles) directions.  Note the
$1/|{\bf r}|^2$ dependence of the interaction strength (as indicated
by the reference line).  Distance is in units of lattice constants,
and the y-axis is in arbitrary units.  (The tail appearing at long
distances is merely an aliasing effect.)
\endinsert

\topinsert\oneandahalfspace
\vskip.5truein
\centerline{\psfig{figure=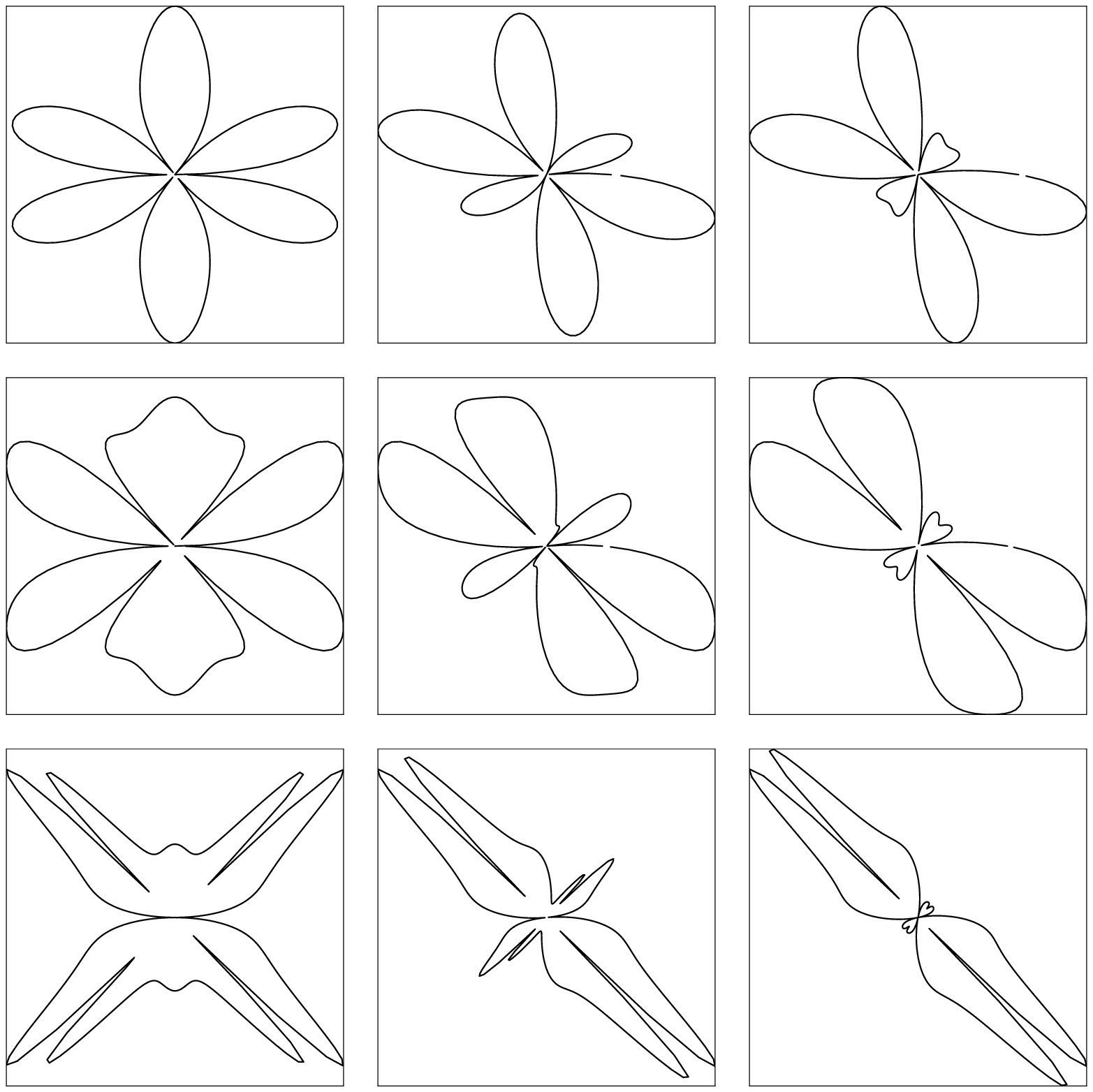,width=4truein}}
\medskip \narrower \parskip 0pt
Figure~11. {\bf Diffraction Contours Due to Linear Coupling to $e_1$: }
Diffraction contour around Bragg peaks $\langle 0 4 \rangle$, $\langle
2 4 \rangle$ and $\langle 2 2 \rangle$ (from left to right) at
an-is-o-tro-pies $\alpha=1$, $\alpha=5$, and $\alpha=50$ (from top to
bottom.)  (The coupling strength $D$ was scaled with the anisotropy,
so that the net deformation is constant in magnitude, but increasingly
tweedy.)  Since tweed can be observed\refto{Muto_Softening} in FePd,
for example, when the anisotropy is as small as $A=5$, this implies
that tweed cannot be explained by a linear response to a coupling to
bulk dilation.
\endinsert

\topinsert\oneandahalfspace
\centerline{\psfig{figure=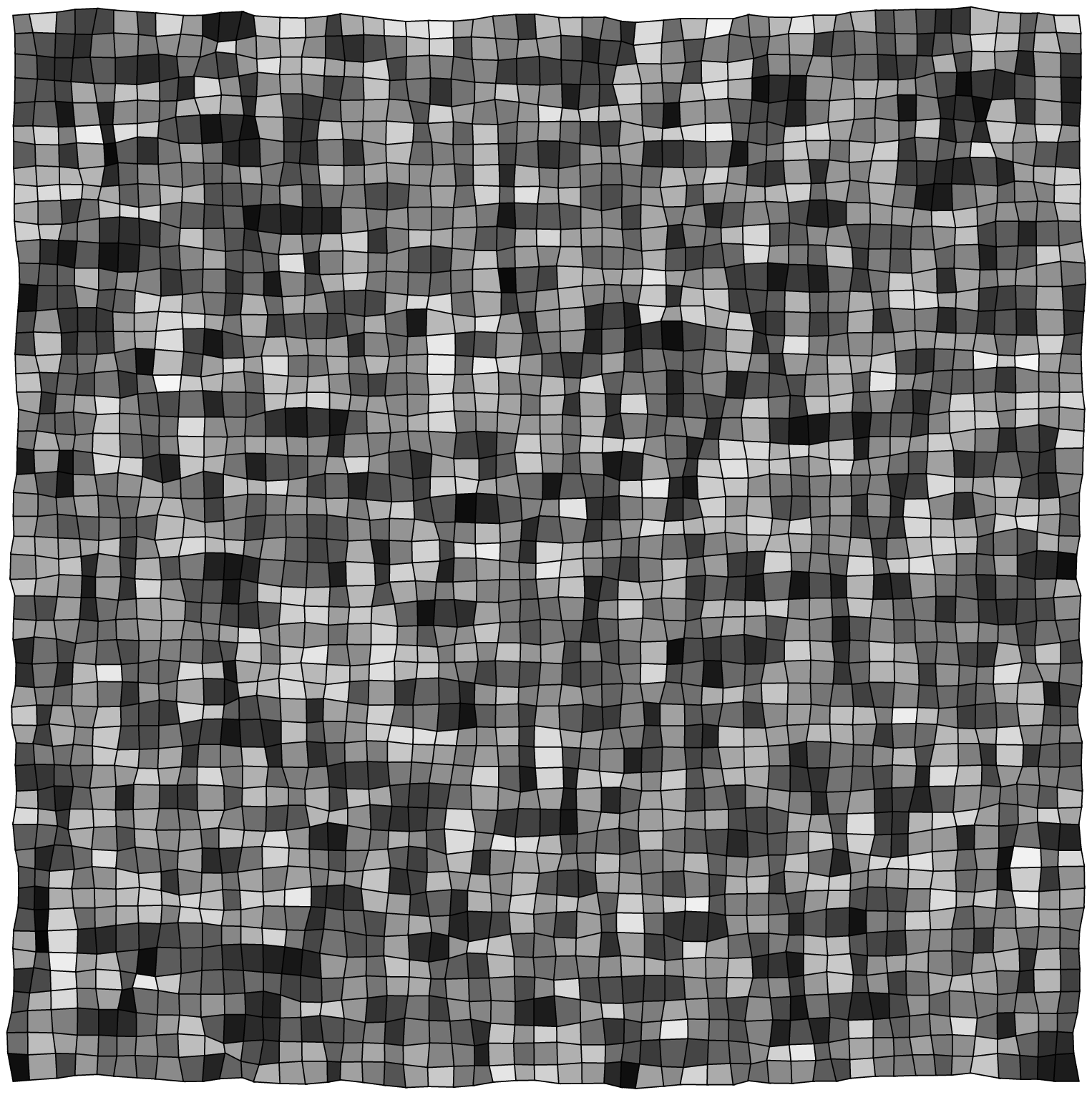,width=4truein} }
\medskip \narrower \parskip 0pt
Figure~12.  {\bf Configuration resulting from linear coupling to
$e_1$:} showing substantial deformation, but no identifiably tweedy
modulation.
\endinsert

\enddocument
\end

\bye